# $\mathbb{Z}_2$ *Topological Index for Continuous Photonic Materials*

*Mário G. Silveirinha*[*]

[(1)]*University of Coimbra, Department of Electrical Engineering – Instituto de Telecomunicações, Portugal, mario.silveirinha@co.it.pt*

**Abstract**

Electronic topological insulators are one of the breakthroughs of the 21st century condensed matter physics. So far, the search for a light counterpart of an electronic topological insulator has remained elusive. This is due to the fundamentally different natures of light and matter and the different spins of photons and electrons. Here, it is shown that the theory of electronic topological insulators has a genuine analogue in the context of light wave propagation in time-reversal invariant continuous materials. We introduce a Gauge invariant $\mathbb{Z}_2$ index that depends on the global properties of the photonic band structure and is robust to any continuous weak variation of the material parameters that preserves the time-reversal invariance. A nontrivial $\mathbb{Z}_2$ index has two possible causes: (i) the lack of smoothness of the pseudo-Hamiltonian in the $\mathbf{k} \to \infty$ limit, and (ii) the entanglement between positive and negative frequency eigenmode branches. In particular, it is proven that electric-type plasmas and magnetic-type plasmas are topologically inequivalent for a fixed wave polarization. We propose a bulk-edge correspondence that links the number of edge modes with the topological invariants of two continuous bulk materials, and present detailed numerical examples that illustrate the application of the theory.

---

[*] To whom correspondence should be addressed: E-mail: *mario.silveirinha@co.it.pt*

# I. Introduction

Topological electronic insulators correspond to a new phase of matter created by the spin-orbit coupling [1-3]. These time-reversal invariant electronic materials are insulating in the bulk, but support an odd number of topologically protected gapless "helical" edge states at the boundary [4, 5]. The number of edge states (modulo 2) is absolutely insensitive to weak perturbations of the material parameters that do not close the gap and maintain the time-reversal invariance [4]. A topologically protected edge state is completely impervious to the presence of imperfections, defects or impurities that do not break the time reversal symmetry. Because of this it has been suggested that topological insulators may be useful for applications in spintronics or in quantum computation.

Triggered by this rather exciting discovery, there have been many attempts in the recent literature to find photonic analogues of electronic topological insulators [6-13]. For example, Khanikaev *et al* proposed a configuration based on matched bianisotropic photonic crystals and that provides an emulation of the Kane-Mele Hamiltonian in a photonic system [6]. A different paradigm relies on a spatially dependent non-periodic optical potential that mimics a synthetic pseudo-magnetic field [10-12]. More recently, it has been shown that (time-reversal invariant) dielectric photonic crystals with the $C_6$ crystal symmetry also provide an interesting form of photonic topological protection [7]. Moreover, even though they are not the focus of the present study, it is relevant to mention that topological effects in nonreciprocal photonic systems, i.e. in systems that do not have the time reversal symmetry, have also been widely discussed in the literature [14-19].

A problem with the designs of periodic photonic topological insulators [6, 7] (which are the counterparts of electronic topological insulators) is that the topological protection that they offer is quite limited [13]. Indeed, while electronic topological insulators are robust to any perturbation that does not violate the time reversal symmetry, the designs of Refs. [6, 7] only provide a topological protection when additional conditions are enforced. Specifically, in Ref. [6] the perturbation must ensure that $\varepsilon = \mu$, whereas in Ref. [7] the perturbation cannot break the $C_6$ crystal symmetry. These are rather harsh restrictions which are ultimately rooted in the fact that the time reversal operator ($\mathcal{T}$) in photonics satisfies $\mathcal{T}^2 = \mathbf{1}$ (because photons are bosons) whereas in electronics $\mathcal{T}^2 = -\mathbf{1}$ (because electrons are fermions) [13]. The property $\mathcal{T}^2 = -\mathbf{1}$ ensures the validity of Kramers theorem, and hence guarantees that at the high-symmetry points of the Brillouin zone the eigenmodes are doubly degenerate [5, 13]. The Kramers theorem is the backbone of the theory of electronic topological insulators, and the lack of an analogue of this theorem in photonics has hindered the progress in the field of topological photonics with the time-reversal symmetry.

In this article, it is demonstrated for the first time that even without the Kramers theorem it is possible to define a $\mathbb{Z}_2$ topological index for continuous bulk photonic materials, which is the counterpart of the $\mathbb{Z}_2$ topological index in electronics. We prove that the $\mathbb{Z}_2$ number is robust to *arbitrary* deformations of the material response that do not break the time-reversal symmetry. We propose a bulk-edge correspondence that links the difference between the $\mathbb{Z}_2$ invariants of two bulk materials with the number of topologically protected edge modes supported by a single interface of the materials. At this point, it is important to highlight that a recent study also suggested that continuous media (specifically hyperbolic chiral media) may support topologically protected edge states [20]. Notably, the theory of Ref. [20] is totally different from ours, and in particular it is based on the calculation of Chern numbers in equifrequency contours (i.e. the surfaces of the form $\omega_\mathbf{k} = const.$ in the wave vector space). Thus, the topological numbers calculated in Ref. [20] do not depend on the global properties of the photonic band structure, but rather on the *single frequency* properties of the equifrequency contours. Rather different, here we show that for continuous media it is possible to introduce a genuine $\mathbb{Z}_2$ index that is related to obstructions to the application of the Stokes theorem in half-wave vector space, and hence depends on the global band structure precisely as in electronics [5, 21]. Hence, our theory establishes for the first time a precise photonic analogue of electronic topological insulators. This paper deals exclusively with continuous media without any granularity. The possibility of extending the theory to photonic crystals is discussed in the final section (Sect. V). The article builds on a previous study on Chern numbers for continuous electromagnetic media [22].

## II. The $\mathbb{Z}_2$ Invariant for Continuous Media

Next, we explore the possibility of defining a $\mathbb{Z}_2$ topological index in photonic bulk materials invariant under the time-reversal ($\mathcal{T}$) operation. We start by reviewing the definitions of the Berry potential and Berry curvature in photonic systems, and then we discuss possible obstructions to the application of the Stokes theorem and how such obstructions may determine a topological invariant in continuous media.

## A. Berry potential and curvature

This work is concerned with uniform continuous lossless media, whose electrodynamics – in absence of field sources – is characterized in the frequency domain by the Maxwell's equations:

$$\hat{N} \cdot \mathbf{F} = \omega \mathbf{G}, \tag{1}$$

where $\mathbf{F} = (\mathbf{E} \ \ \mathbf{H})^T$, $\mathbf{G} = (\mathbf{D} \ \ \mathbf{B})^T$, are six-component vector fields ($T$ denotes the transpose of a vector), $\mathbf{E}, \mathbf{H}$ are the electric and magnetic fields, $\mathbf{D}, \mathbf{B}$ are the electric displacement and the induction fields, and $\omega$ is the oscillation frequency. In the above,

$$\hat{N}(-i\nabla) = \begin{pmatrix} \mathbf{0} & i\nabla \times \mathbf{1}_{3\times 3} \\ -i\nabla \times \mathbf{1}_{3\times 3} & \mathbf{0} \end{pmatrix}$$

is a differential operator, being $\mathbf{1}_{3\times 3}$ the identity tensor of dimension three and $\nabla = \frac{\partial}{\partial x}\hat{\mathbf{x}} + \frac{\partial}{\partial y}\hat{\mathbf{y}} + \frac{\partial}{\partial z}\hat{\mathbf{z}}$. We admit that $\mathbf{F}$ and $\mathbf{G}$ are linked in the spectral domain by a general bianisotropic constitutive relation [23, 24]:

$$\mathbf{G} = \mathbf{M} \cdot \mathbf{F}, \quad \text{with} \quad \mathbf{M}(\omega) = \begin{pmatrix} \varepsilon_0 \overline{\varepsilon} & \frac{1}{c}\overline{\xi} \\ \frac{1}{c}\overline{\zeta} & \mu_0 \overline{\mu} \end{pmatrix}. \tag{2}$$

where the tensors $\overline{\varepsilon}(\omega), \overline{\mu}(\omega), \overline{\xi}(\omega), \overline{\zeta}(\omega)$ are dimensionless and represent the frequency-dependent permittivity, permeability and the magneto-electric coupling tensors, respectively.

Let us consider some family of eigenmodes $\mathbf{F}_{n\mathbf{k}} = \mathbf{f}_{n\mathbf{k}} e^{i\mathbf{k} \cdot \mathbf{r}}$ with envelope $\mathbf{f}_{n\mathbf{k}}$ associated with the eigenfrequencies $\omega_{n\mathbf{k}}$. The index $n$ identifies the eigenmode branches. It is assumed without loss of generality that the wave vector $\mathbf{k}$ is of the form $\mathbf{k} = k_x \hat{\mathbf{x}} + k_y \hat{\mathbf{y}}$, i.e. we are interested in the wave propagation in the *xoy* plane. Other planes of propagation can be treated in a similar way. The $\mathbb{Z}_2$ topological invariant may depend on the considered plane of propagation. In continuous media the eigenmodes are necessarily plane waves ($\mathbf{f}_{n\mathbf{k}}$ is independent of $\mathbf{r}$), and thus the electromagnetic field envelope satisfies $\left[\hat{N}(\mathbf{k}) - \omega_{n\mathbf{k}} \mathbf{M}(\omega_{n\mathbf{k}})\right] \cdot \mathbf{f}_{n\mathbf{k}} = 0$. It was originally shown by Raghu and Haldane that the Berry potential of non-bianisotropic local media may be written in terms of $\mathbf{f}_{n\mathbf{k}}$ as [14, 15]:

$$\mathcal{A}_{n\mathbf{k}} = \frac{\mathrm{Re}\left\{ i \mathbf{f}_{n\mathbf{k}}^* \cdot \frac{\partial}{\partial \omega}\left[\omega \mathbf{M}(\omega)\right]_{\omega_{n\mathbf{k}}} \cdot \partial_{\mathbf{k}} \mathbf{f}_{n\mathbf{k}} \right\}}{\mathbf{f}_{n\mathbf{k}}^* \cdot \frac{\partial}{\partial \omega}\left[\omega \mathbf{M}(\omega)\right]_{\omega_{n\mathbf{k}}} \cdot \mathbf{f}_{n\mathbf{k}}}. \qquad (3)$$

where $\partial_{\mathbf{k}} = \frac{\partial}{\partial k_x}\hat{\mathbf{x}} + \frac{\partial}{\partial k_y}\hat{\mathbf{y}}$. Recently, we demonstrated in Ref. [22] that the above formula also holds for bianisotropic media and for a subclass of spatially dispersive media. The material matrix needs to satisfy the restriction $\frac{\partial}{\partial \omega}\left[\omega \mathbf{M}(\omega)\right] > 0$, that is $\frac{\partial}{\partial \omega}\left[\omega \mathbf{M}(\omega)\right]$ must be positive definite [14, 15, 22]. The Berry curvature $\mathcal{F}_{\mathbf{k}}$ is written in terms of the Berry potential as:

$$\mathcal{F}_{\mathbf{k}} = \frac{\partial \mathcal{A}_y}{\partial k_x} - \frac{\partial \mathcal{A}_x}{\partial k_y}. \qquad (4)$$

where it is implicit that $\mathcal{A}_{\mathbf{k}}$ includes the contributions from all the relevant photonic bands.

We are interested in materials with a response invariant under the time-reversal operator $\mathcal{T}$. The time-reversal operator for photonic systems is of the form $\mathcal{T} = \mathcal{K}\mathcal{U}$ where $\mathcal{K}$ denotes the complex conjugation operator and $\mathcal{U} = \begin{pmatrix} \mathbf{1}_{3\times 3} & 0 \\ 0 & -\mathbf{1}_{3\times 3} \end{pmatrix}$. For time-reversal invariant systems $\mathcal{T}: \mathbf{f}_{n\mathbf{k}} \to \mathcal{T} \cdot \mathbf{f}_{n\mathbf{k}}$ maps eigenmodes $\mathbf{f}_{n\mathbf{k}}$ associated with the frequency and wave vector pair $(\omega_{n\mathbf{k}}, \mathbf{k})$ into eigenmodes $(\mathcal{T} \cdot \mathbf{f}_{n\mathbf{k}})$ associated with the pair $(\omega_{n\mathbf{k}}, -\mathbf{k})$, i.e. $\mathcal{T}$ flips the wave vector.

The time reversal invariance requires that $\mathbf{M}(\omega) = \mathcal{U} \cdot \mathbf{M}^*(\omega) \cdot \mathcal{U}$. For lossless systems (for which $\mathbf{M}(\omega)$ is Hermitian symmetric for $\omega$ real-valued [24]) the time-reversal invariance is equivalent to the Lorentz reciprocity. The constitutive parameters of a reciprocal material (e.g. standard dielectrics and metals) are required to satisfy [24]:

$$\overline{\varepsilon}(\omega) = \left[\overline{\varepsilon}(\omega)\right]^T, \qquad \overline{\mu}(\omega) = \left[\overline{\mu}(\omega)\right]^T, \qquad \overline{\zeta}(\omega) = -\left[\overline{\xi}(\omega)\right]^T, \qquad (5)$$

where the superscript "$T$" denotes the transpose matrix. For a system with the $\mathcal{T}$ symmetry the Berry curvature satisfies $\mathcal{F}_{\mathbf{k}} = -\mathcal{F}_{-\mathbf{k}}$, and hence the corresponding Chern numbers vanish [13, 15].

As discussed in Ref. [22], for continuous media it is useful to visualize the wave vector space ($k_x, k_y \in [-\infty, +\infty]$) as the Riemann sphere (Fig. 1a). Each point of the $(k_x, k_y)$ plane can be mapped into a point $\boldsymbol{\kappa}$ of the Riemann sphere by the stereographic projection. The origin is mapped into the south-pole (*S*), and the point $\mathbf{k} = \infty$ is mapped into the north-pole (*N*). In the rest of the article, we loosely identify the points of the plane with the points of the Riemann sphere whenever it is pertinent.

In electronic systems the $\mathbb{Z}_2$ topological invariant can be calculated by counting the zeros of a Pfaffian function [1, 5, 21]. The topological invariant may also be expressed in terms of integrals involving the Berry potential, such that the integration region consists of half-Brillouin zone, i.e. *half* of the wave vector space [5, 21]. It will be shown in this article that it is possible to extend such a concept to continuous photonic systems. Thus, we define the effective Brillouin zone ( *EBZ* ) as half of the Riemann sphere surface (e.g. the half of the sphere in the semi-plane $k_y \geq 0$), and the $\partial EBZ$ (a meridian circle) as the corresponding boundary (Fig. 1a).

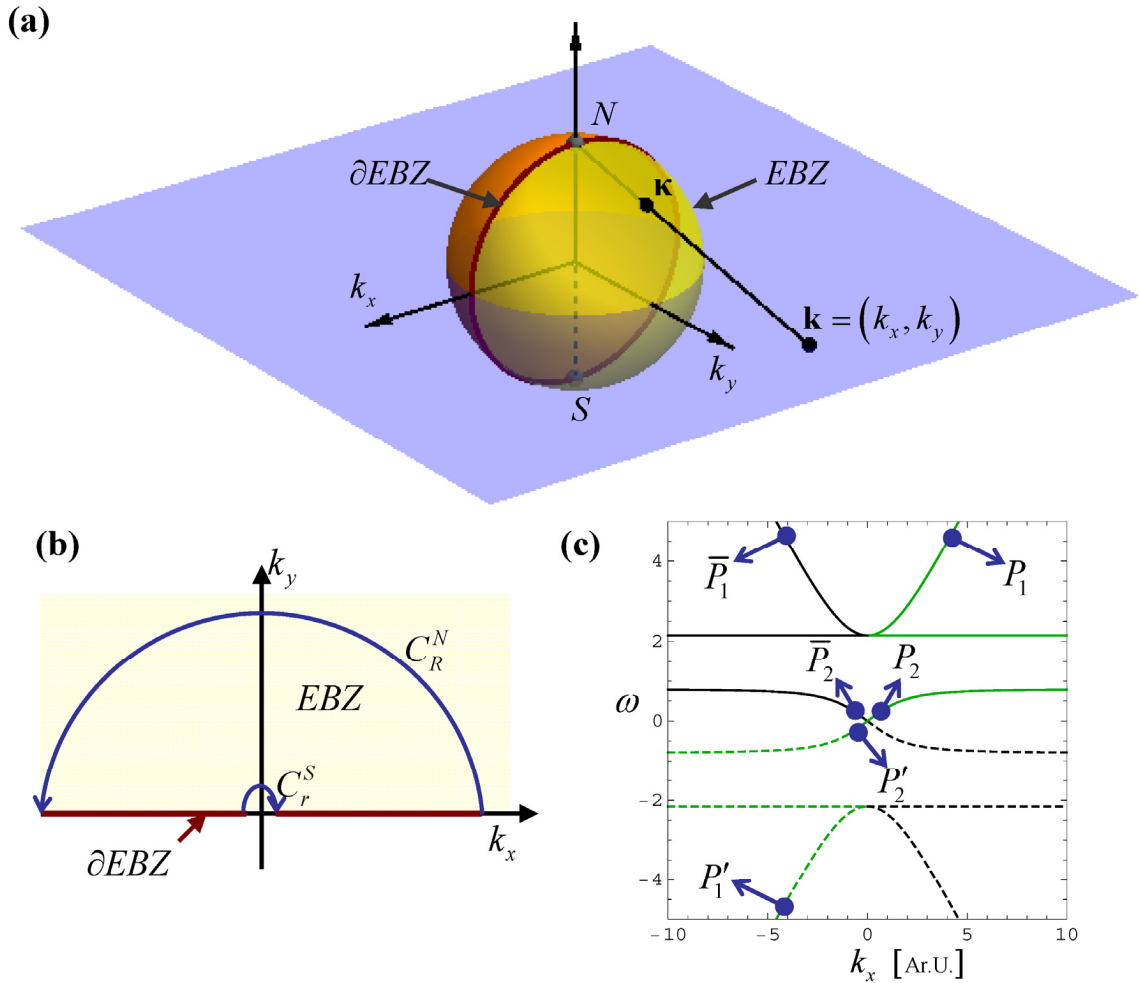

Fig. 1. (Color online) **(a)** The wave vector space can be regarded as the Riemann sphere. Each point of the sphere is projected into a single point of the plane through the stereographic projection. Half of the wave vector space (e.g. the semi-space $k_y > 0$) is mapped into half of the Riemann sphere surface: the effective Brillouin Zone (*EBZ*). The boundary of the *EBZ* is represented in the Riemann sphere by a circle ($\partial EBZ$).

**(b)** Integration contours used in Eq. (11). **(c)** Generic band diagram of a lossless material with the time-reversal and reality symmetries, showing both the positive and negative parts of the frequency spectrum.

Because the electromagnetic fields are real-valued physical entities (different from the wave function in electronics), the material matrix is required to satisfy as well the "reality condition" [25], i.e. $\mathbf{M}(\omega) = \mathbf{M}^*(-\omega)$ for $\omega$ real valued. In other words, the Maxwell's equations are invariant under the application of the complex conjugation operator $\mathcal{K}$. As a consequence, it is possible to transform a natural mode associated with the pair $(\omega_{n\mathbf{k}}, \mathbf{k})$ into another eigenmode associated with the pair $(-\omega_{n\mathbf{k}}, -\mathbf{k})$ through the mapping $\mathcal{K}: \mathbf{f}_{n\mathbf{k}} \to \mathcal{K} \cdot \mathbf{f}_{n\mathbf{k}}$. Hence, as illustrated in Fig. 1c, the frequency spectrum consists of both positive and negative frequency branches. Evidently, the positive and negative frequency branches are generally disconnected from each other, except in the long wavelength limit $(\omega, \mathbf{k}) \approx (0, \mathbf{0})$ (see points $P_2$ and $P_2'$ in Fig. 1c) and at infinity $(\omega, \mathbf{k}) \approx (\infty, \infty)$ (see points $P_1$ and $P_1'$). It will be shown later that the connections of the positive and negative frequency spectra play an important role in the definition of the $\mathbb{Z}_2$ number for photonic continuous media.

### B. *Obstructions to the application of the Stokes theorem*

Let us consider a globally defined Gauge of eigenmodes $\mathbf{f}_{n\mathbf{k}}$ that varies smoothly in all space, except possibly at a few isolated points. Usually, the singularities occur at the points of the wave vector space wherein the group velocity of the electromagnetic modes vanishes. In a periodic system with the time-reversal symmetry ($\omega_{n\mathbf{k}} = \omega_{n,-\mathbf{k}}$) the group velocity usually vanishes at all the high-symmetry points of the Brillouin zone that stay invariant under the action of $\mathcal{T}$ (e.g. for a square lattice these points are $\Gamma, X, Y,$ and $M$ following the usual notations). In a continuous system, the points invariant under $\mathcal{T}$ are the $\mathbf{k} = \infty$ and $\mathbf{k} = 0$ points, i.e. the $N$ and $S$ poles of the Riemann sphere.

The reason why a zero-group velocity may imply a singular behavior of the individual eigenwave branches is because these points behave as either "sources" or "sinks" of the Poynting vector lines, and such a property may be incompatible with having both $\mathbf{E}_\mathbf{k}$ and $\mathbf{H}_\mathbf{k}$ continuous. This is best illustrated with an example. Consider a lossless electron gas described by the Drude model $\varepsilon = 1 - \omega_{pe}^2 / \omega^2$ and $\mu = 1$ where $\omega_{pe}$ is the plasma frequency. For propagation in the $xoy$ plane there are three families of positive frequency electromagnetic modes: the transverse magnetic (TM) and the transverse electric (TE) waves and the longitudinal electric waves (LE) such that

$$\mathbf{f}_{\mathrm{TM},\mathbf{k}} \sim \begin{pmatrix} -\dfrac{\mathbf{k}}{\omega_{\mathrm{T},\mathbf{k}} \varepsilon_0 \varepsilon(\omega_{\mathrm{T},\mathbf{k}})} \times \hat{\mathbf{z}} \\ \hat{\mathbf{z}} \end{pmatrix}, \qquad \mathbf{f}_{\mathrm{TE},\mathbf{k}} \sim \begin{pmatrix} \hat{\mathbf{z}} \\ \dfrac{\mathbf{k}}{\omega_{\mathrm{T},\mathbf{k}} \mu_0 \mu} \times \hat{\mathbf{z}} \end{pmatrix}, \qquad \mathbf{f}_{\mathrm{LE},\mathbf{k}} \sim \begin{pmatrix} \mathbf{k} \\ 0 \end{pmatrix}. \qquad (6)$$

In the above, $\omega_{\mathrm{T},\mathbf{k}} = \sqrt{\omega_{pe}^2 + k^2 c^2}$ is the dispersion of the transverse eigenmodes, and the normalization of the modes is arbitrary (typically the modes are normalized to have unit energy [22]). The longitudinal modes are dispersionless ($\omega_{\mathrm{L},\mathbf{k}} = \omega_{pe}$). Notably, for TM waves the magnetic field is always directed along $z$ and hence to have an outward radial Poynting vector ("source"-type lines), the electric field is forced to wind around the origin. As a consequence $\mathbf{f}_{\mathrm{TM},\mathbf{k}}$ is discontinuous in the vicinity of the origin (note that when $k \to 0$ one has $\mathbf{f}_{\mathrm{TM},\mathbf{k}} \sim \begin{pmatrix} \hat{\mathbf{k}} \times \hat{\mathbf{z}} \\ 0 \end{pmatrix}$ i.e. the magnetic field becomes negligible compared to the electric field). The longitudinal mode is also discontinuous at the origin. Importantly, it is impossible to get rid of the discontinuous behavior of each branch near the origin with a Gauge transformation. It is underlined that the time reversal invariance forces the Poynting vector lines to have at least a "source" and a "sink" in the wave vector space.

The singularities of the eigenmodes $\mathbf{f}_{n\mathbf{k}}$ at the time-reversal invariant points of the wave vector space may lead to an obstruction to the application of the Stokes theorem to the Berry potential. Our motivation to study possible obstructions to the Stokes theorem is inspired by the findings of Fu and Kane (Ref. [21]) for electronic systems [5].

To show this, let us consider first an arbitrary closed contour $C$ in the $\mathbf{k}$-plane such that $\mathbf{f}_{n\mathbf{k}}$ is free of singularities on the contour $C$. However, the individual eigenmode branches $\mathbf{f}_{n\mathbf{k}}$ may be singular at a few points $\mathbf{k}_m^s$ ($m$=1,2,..) interior to $C$. Let us define the quantity

$$\mathcal{D}_C = \frac{1}{2\pi}\left(\oint_C \mathcal{A}_\mathbf{k} \cdot d\mathbf{l} - \int_{\text{int}\,C} \mathcal{F}_\mathbf{k} ds\right), \tag{7}$$

where $\mathcal{A}_\mathbf{k}$ and $\mathcal{F}_\mathbf{k}$ are the Berry potential and curvature associated with the considered set of eigenfunctions. The first integral is a line integral, and the second integral is a surface integral over the region interior to $C$ ($\text{int}\,C$). The direct application of the Stokes theorem to $C$ would give $\mathcal{D}_C = 0$. Evidently, the singularity points $\mathbf{k}_m^s$ may be an obstruction to the result $\mathcal{D}_C = 0$. Indeed, it is clear that in general one has:

$$\mathcal{D}_C = \sum_m \frac{1}{2\pi} \oint_{C_r(\mathbf{k}_m^s)} \mathcal{A}_\mathbf{k} \cdot d\mathbf{l}. \tag{8}$$

In the above, $C_r(\mathbf{k}_m^s)$ represents a circle centered at $\mathbf{k}_m^s$ with radius $r \to 0^+$. It can be shown (see Ref. [22] for detailed arguments) that because the relevant pseudo-Hamiltonian is smooth for any finite $\mathbf{k}$ the integral $\frac{1}{2\pi}\oint_{C_r(\mathbf{k}_m^s)}\mathcal{A}_\mathbf{k}\cdot d\mathbf{l}$ is necessarily some integer number $l_m$. Thus, in general $\mathcal{D}_C$ is a nonzero integer[†].

So far our discussion is completely general. Let us now consider the interesting case wherein the material has the time reversal symmetry and $C$ is the effective Brillouin zone boundary ($\partial EBZ$), so that the interior of $C$ is exactly half of the wave vector space ($EBZ$). Specifically, if the point $\mathbf{k}$ belongs to the $EBZ$ then the point $-\mathbf{k}$ cannot belong to the $EBZ$, except if it lies on the boundary $\partial EBZ$. Figure 1a depicts the particular case wherein the $EBZ$ is the semi-plane $k_y \geq 0$, but other choices are allowed. Crucially, independent of the choice of the $EBZ$ the points $\mathbf{k}=\infty$ and $\mathbf{k}=\mathbf{0}$ (north and south poles of the Riemann sphere) are always in the $\partial EBZ$ because they are invariant under the time-reversal transformation. Therefore, it is always feasible to choose the $EBZ$ in such a manner that there are no singularities $\mathbf{k}_m^s$ on the boundary $\partial EBZ$, with the possible exceptions of the north and south poles of the Riemann sphere. Thus, compared to the case discussed previously, now there is an additional cause for an obstruction to the Stokes theorem: the possible singularities at the time-reversal invariant points. Let us write,

$$\mathcal{D}_{EBZ} = \frac{1}{2\pi}\left(\oint_{\partial EBZ}\mathcal{A}_\mathbf{k}\cdot d\mathbf{l} - \int_{EBZ}\mathcal{F}_\mathbf{k}ds\right). \tag{9}$$

---

[†] In some cases (for bounded contours $C$) it may be possible to make a Gauge transformation such that the transformed basis remains globally defined and has $\mathcal{D}_C = 0$.

For example, if the $EBZ$ is taken as the semi-plane $k_y \geq 0$ the integrals can be explicitly spelled out as $\mathcal{D}_{EBZ} = \frac{1}{2\pi}\left(\int_{-\infty}^{+\infty} \mathcal{A}_\mathbf{k} \cdot \hat{\mathbf{x}}\, dk_x - \iint_{k_y>0} \mathcal{F}_\mathbf{k}\, dk_x dk_y\right)$. When the eigenfunctions are discontinuous at the $N$ and $S$ poles, the Berry potential can also be discontinuous at these points. Hence, it is convenient to make the definition of $\mathcal{D}_{EBZ}$ more precise:

$$\mathcal{D}_{EBZ} = \frac{1}{2\pi} \lim_{\substack{r \to 0^+ \\ R \to \infty}} \left[ \left(\int_{-R}^{-r} + \int_{r}^{R}\right) \mathcal{A}_\mathbf{k} \cdot \hat{\mathbf{x}}\, dk_x - \iint_{\substack{k_y > 0 \\ r \leq |\mathbf{k}| \leq R}} \mathcal{F}_\mathbf{k}\, dk_x dk_y \right]. \tag{10}$$

Using the Stokes theorem in a domain where the singularities are excluded one finds that,

$$\mathcal{D}_{EBZ} = \frac{-1}{2\pi} \int_{C_r^S} \mathcal{A}_\mathbf{k} \cdot d\mathbf{l} + \frac{-1}{2\pi} \int_{C_R^N} \mathcal{A}_\mathbf{k} \cdot d\mathbf{l} + \frac{1}{2\pi} \sum_{\mathbf{k}_m^s \in EBZ} \oint_{C_r(\mathbf{k}_m^s)} \mathcal{A}_\mathbf{k} \cdot d\mathbf{l}, \tag{11}$$

where $C_r^S$ and $C_R^N$ are *half-circles* centered respectively at the $S$ and $N$ poles, which in the $\mathbf{k}$-plane correspond to the half-circles shown in the Fig. 1b. It is implicit in the above formula that $r \to 0^+$ and $R \to \infty$. Note that in the Riemann sphere both $C_r^S$ and $C_R^N$ have vanishingly small radius when $r \to 0^+$ and $R \to \infty$, respectively. Thus, $\mathcal{D}_{EBZ}$ has contributions from the possible singularities at the $S$ and $N$ poles (first two terms of Eq. (11)), and from the singularities interior to the $EBZ$ (third term of Eq. (11) which was previously shown to be an integer number).

## C.    *Gauge constraints*

Next, we discuss the possibility of imposing some Gauge restrictions on the globally defined basis of eigenfunctions. The Gauge restrictions are crucial to define a topological number. Specifically, let $\mathbf{f}_{n\mathbf{k}}$ be some globally defined family of eigenmode branches that includes all the Bloch waves in some frequency range of the form $\omega_{min} < \omega < \omega_{max}$ where $\omega_{min}, \omega_{max}$ should be nonnegative frequencies in a bandgap or alternatively $\omega_{min} = 0$ or $\omega_{max} = \infty$. Note that only *positive* frequencies are included in the range of interest.

As discussed in the previous subsection, in general the individual branches $\mathbf{f}_{n\mathbf{k}}$ are not continuous at the *S* and *N* poles. Nevertheless, the eigenspaces of the pseudo-Hamiltonian vary continuously with the wave vector for any $\mathbf{k}$-finite because the pseudo-Hamiltonian is a smooth function of $\mathbf{k}$ [22]. In particular, the eigenspaces vary continuously in the vicinity of south-pole ($\mathbf{k} = 0$). This implies that $(\mathbf{f}_{n\mathbf{k}})_{S^-}$ and $(\mathbf{f}_{n\mathbf{k}})_{S^+}$ are related by some unitary transformation $\mathbf{V}$, where the $S^\pm$ points belong to the $\partial EBZ$ and are displaced by an infinitesimal amount to the right/left of the *S* pole. For example, if the $EBZ$ is the region $k_y \geq 0$ the points $S^\pm$ correspond in the **k**-plane to $(0^\pm, 0)$. Thus, it is always possible to pick a Gauge such that:

$$(\mathbf{f}_{n\mathbf{k}})_{S^-} = (\mathbf{f}_{n\mathbf{k}})_{S^+}, \qquad \text{for the branches with } \omega_{n,k=0^+} \neq 0. \tag{12a}$$

The identity $(\mathbf{f}_{n\mathbf{k}})_{S^-} = (\mathbf{f}_{n\mathbf{k}})_{S^+}$ should be understood as: *the basis* $(\mathbf{f}_{n\mathbf{k}})_{S^-}$ *is linked with the basis* $(\mathbf{f}_{n\mathbf{k}})_{S^-}$ *by some unitary transformation* $\mathbf{V}$ *with* $\det(\mathbf{V}) = 1$. Often $\mathbf{V}$ can be taken as the identity matrix, and in that case Eq. (12a) is equivalent to the continuity of the individual branches at the *S* point with the wave vector restricted to the $\partial EBZ$.

Note that the positive frequency eigenmode branches with $\omega_{n,k=0^+} = 0^+$ are not included in Eq. (12a). The reason is that the $\omega = 0$ eigenspace is reached both as the limit of positive and negative frequency eigenspaces (Fig. 1c). Hence, it is generally impossible to link the $\omega = 0$ eigenspaces at the $S^\pm$ points by a unitary transformation without considering both the positive and the negative frequency branches (and also the zero-frequency dispersionless branches, if any). For example, in the long wavelength limit ($\omega = 0^+$) the eigenfunctions usually are alike the eigenfunctions of the vacuum. Notably, the Poynting vector of the electromagnetic modes does not vanish in the $\omega = 0^+$ limit. Thus, in the long wavelength limit the point $S$ ($\mathbf{k} = 0$) behaves as a "source" of the Poynting vector lines of the relevant positive frequency branches $\mathbf{f}_{n\mathbf{k}}$. In particular, the Poynting vector changes direction at points of the form $S^-$ and $S^+$ (in Fig. 1c these points may be identified with $\bar{P}_2$ and $P_2$). Thus, $\mathbf{f}_{n\mathbf{k}}$ cannot possibly be the same at $S^-$ and $S^+$ because this would imply the continuity of the Poynting vector. This confirms that the eigenspaces generated by the positive frequency eigenfunctions with $\omega_{n,k=0^+} = 0^+$ are indeed different at the $S^\pm$ points.

Because of the time reversal symmetry, the $\omega = 0^+$ eigenspaces at the $S^\pm$ points can always be linked by the time reversal operator $\mathcal{T}$. Importantly, in the long wavelength limit $\omega \to 0^+$ the magneto-electric coupling parameters vanish, and hence the material response $\mathbf{M}(\omega = 0^+)$ is real-valued. This means that the $\omega = 0^+$ eigenspaces can be generated by real-valued vectors, and hence these eigenspaces are invariant under the application of the complex conjugation operator $\mathcal{K}$. This discussion shows that one can connect the $\omega = 0^+$ eigenspaces at the $S^\pm$ points through the operator $\mathcal{U} = \mathcal{K}\mathcal{T}$. Thus, we can pick a Gauge such that:

$$\left(\mathbf{f}_{n\mathbf{k}}\right)_{S^-} = \left(\mathcal{U} \cdot \mathbf{f}_{n\mathbf{k}}\right)_{S^+}, \text{ for the branches with } \omega_{n,k=0^+} = 0. \tag{12b}$$

As before, the identity $(\mathbf{f}_{n\mathbf{k}})_{S^-} = (\mathcal{U} \cdot \mathbf{f}_{n\mathbf{k}})_{S^+}$ means that the two eigenfunctions basis must be related by a unitary transformation $\mathbf{V}$ with $\det(\mathbf{V}) = 1$.

We want to impose that the eigenfunctions basis satisfies constraints analogous to (12a) and (12b) at the north-pole ($k = \infty$). Specifically, we want to pick a Gauge for which:

$$(\mathbf{f}_{n\mathbf{k}})_{N^-} = (\mathbf{f}_{n\mathbf{k}})_{N^+}, \qquad \text{for the branches with } \omega_{n,k=\infty} \neq \infty. \tag{12c}$$

$$(\mathbf{f}_{n\mathbf{k}})_{N^-} = (\mathcal{U} \cdot \mathbf{f}_{n\mathbf{k}})_{N^+}, \text{ for the branches with } \omega_{n,k=\infty} = \infty. \tag{12d}$$

Here, $N^-$ and $N^+$ represent points in the $\partial EBZ$ offset by an infinitesimal amount from the $N$ point.

The justification for Eq. (12d) is analogous to that given in Eq. (12b). Indeed, in the high-frequency limit ($\omega = \infty$) the branches of positive and negative frequencies are effectively entangled at the north-pole. Thus, the eigenspaces generated by $(\mathbf{f}_{n\mathbf{k}})$ at the $N^-$ and $N^+$ points are usually different (in Fig. 1c these points are $\bar{P}_1$ and $P_1$). Indeed, for $\omega = \infty^+$ the point $N$ effectively behaves as a "sink" of the Poynting vector lines. Crucially, in the high-frequency limit the material response should be asymptotically analogous to that of the vacuum and hence the magneto-electric coupling response is required to vanish. In other words, $\mathbf{M}(\omega = \infty)$ must be real-valued. Hence, the $\omega = +\infty$ eigenspaces at the $N^\pm$ points can be connected by the operator $\mathcal{U} = \mathcal{KT}$, consistent with Eq. (12d).

Regarding the constraint in Eq. (12c), it can be enforced when the eigenspaces at the $N^\pm$ points are the same for a finite eigenvalue $\omega_{n,k=\infty}$. This is guaranteed to happen if the pseudo-Hamiltonian of the system varies smoothly at the north pole of the Riemann sphere. The interesting thing is that for continuous media the pseudo-Hamiltonian is usually ill-behaved at the north-pole [22], and hence it is not obvious *a priori* that the eigenspaces at the $N^{\pm}$ need to be same. In Appendix A, it is proven that notwithstanding that the pseudo-Hamiltonian is not smooth at infinity it is always possible to impose the constraint (12c).

## D. The $\mathbb{Z}_2$ invariant

We are now ready to define a $\mathbb{Z}_2$ index for continuous photonic systems with the time-reversal symmetry. Let us then consider a globally defined basis of eigenfunctions $\mathbf{f}_{n\mathbf{k}}$ associated with the eigenvalues lying in some frequency range $\omega_{min} < \omega < \omega_{max}$ (with $0 \leq \omega_{min}, \omega_{max} \leq \infty$), such that the basis $\mathbf{f}_{n\mathbf{k}}$ satisfies the Gauge constraints discussed in the previous subsection [Eq. (12)]. We define the $\mathbb{Z}_2$ number as $\mathcal{D} = (2\mathcal{D}_{EBZ}) \mod 2$ so that:

$$\mathcal{D} = \frac{1}{\pi}\left(\oint_{\partial EBZ} \mathcal{A}_{\mathbf{k}} \cdot \mathbf{dl} - \int_{EBZ} \mathcal{F}_{\mathbf{k}} ds\right) \mod 2. \tag{13}$$

It will be shown in what follows that $\mathcal{D}$ is a Gauge invariant integer number. From the subsection II.B, it is clear that a nontrivial topological number $\mathcal{D}$ may be regarded as an obstruction to the application of the Stokes theorem to half-wave vector space when the picked Gauge satisfies the constraints (12), somewhat similar to what happens in electronic systems [5, 21]. The definition of the $\mathbb{Z}_2$ number in Eq. (13) differs by a factor of two from that proposed by Fu and Kane for electronic systems $\tilde{\mathcal{D}} = \mathcal{D}_{EBZ} \mod 2$ [21]. It is shown in Appendix B that the two definitions can be reconciled if the Berry potential includes the contribution from both positive and negative frequency eigenfunctions.

To begin with, we observe that from Eq. (11) $\mathcal{D}$ may have contributions from the singularities interior to the *EBZ* and from the time reversal invariant points *N* and *S*. From section II.B it is clear that $2 \times \frac{1}{2\pi} \sum_{\mathbf{k}_m^s \in EBZ} \oint_{C_r(\mathbf{k}_m^s)} \mathcal{A}_\mathbf{k} \cdot \mathbf{dl} = 2 l_m$ is an even integer number, and hence the singularities in the interior of the *EBZ* do not contribute to $\mathcal{D}$. Therefore, Eq. (11) implies that the $\mathcal{D}$ number can be written as:

$$\mathcal{D} = \left[ \frac{-1}{\pi} \int_{C_r^S \cup C_R^N} \mathcal{A}_\mathbf{k} \cdot \mathbf{dl} \right] \mod 2, \tag{14}$$

i.e., the invariant can be calculated simply by integrating the globally defined and Gauge constrained Berry potential over semicircles with infinitesimal radius that encircle the *N* and *S* points of the Riemann sphere.

The $\mathcal{D}$ number is Gauge invariant when the globally defined basis is restricted to satisfy the constraints (12). The Gauge invariance can be proven by noting that a smooth Gauge transformation of the positive frequency branches, $\mathbf{f}_{n\mathbf{k}} \to \mathbf{f}_{n\mathbf{k}} e^{i\theta_{n\mathbf{k}}}$, must be such that

$$(\theta_{n\mathbf{k}})_{N^-} - (\theta_{n\mathbf{k}})_{N^+} = 2\pi l_n \quad \text{and} \quad (\theta_{n\mathbf{k}})_{S^-} - (\theta_{n\mathbf{k}})_{S^+} = 2\pi l_n'. \tag{15}$$

where $l_n, l_n'$ are integers. Indeed, only in these conditions the new Gauge satisfies Eqs. (12). Because the Berry potential is transformed as $\mathcal{A}_\mathbf{k} \to \mathcal{A}_\mathbf{k} - \nabla_\mathbf{k} \sum_n \theta_{n\mathbf{k}}$ [5], it follows from Eq. (14) that the $\mathcal{D}$ number is transformed as

$$\mathcal{D} \to \left[ \frac{-1}{\pi} \int_{C_r^S \cup C_R^N} \mathcal{A}_\mathbf{k} \cdot \mathbf{dl} + \frac{1}{\pi} \int_{C_r^S \cup C_R^N} \nabla_\mathbf{k} \theta_{n\mathbf{k}} \cdot \mathbf{dl} \right] \mod 2, \text{ or equivalently:}$$

$$\mathcal{D} \to \left[ \frac{-1}{\pi} \int_{C_r^S \cup C_R^N} \mathcal{A}_\mathbf{k} \cdot d\mathbf{l} + \frac{1}{\pi} \left( \theta_{n,\mathbf{k}=0^+} - \theta_{n,\mathbf{k}=0^-} \right) + \frac{1}{\pi} \left( \theta_{n,\mathbf{k}=-\infty} - \theta_{n,\mathbf{k}=+\infty} \right) \right] \mod 2 = \mathcal{D}, \qquad (16)$$

where the last identity is a consequence of the constraints in (15) and it is implicit that $\mathbf{k}$ is in the $\partial EBZ$. This proves that $\mathcal{D}$ is really Gauge invariant.

Furthermore, it is proven in Appendix C that $\mathcal{D}$ is a $\mathbb{Z}_2$ integer. Hence, bulk continuous photonic media can be classified in a given spectral range as topologically trivial ($\mathcal{D}=0$) or topologically nontrivial ($\mathcal{D}=1$). For continuous media with $\mathcal{D}=1$ it is impossible to directly apply the Stokes theorem to the $EBZ$ when the eigenfunctions basis is required to satisfy the restrictions in Eq. (12). Moreover, the analysis of Appendix C reveals that the nontrivial contributions to the topological invariant have two origins: (*i*) the singular nature of the pseudo-Hamiltonian at the $N$ pole ($k=\infty$). (*ii*) the entanglement between positive and negative eigenmode branches at either $\omega=0^+$ or $\omega=+\infty$. In particular, the integral of the Berry potential over the half-circle $C_r^S$ gives a trivial contribution to the invariant, with the exception of branches with $\omega_{n,k=0} = 0^+$.

To conclude this section, we note that the integral in Eq. (14) (over two half-circles of infinitesimal radius in the Riemann sphere) in principle varies continuously if the $EBZ$ is smoothly deformed (so that the points $N^\pm$ and $S^\pm$ are also varied continuously), ensuring at the same time that the Gauge restrictions (12) are satisfied (thus, the picked Gauge must also be smoothly deformed). But since $\mathcal{D}$ is an integer, this indicates that it stays invariant under a smooth deformation of the $EBZ$, and hence $\mathcal{D}$ is independent of the specific choice of the $EBZ$.

# III. Isotropic Dielectrics

To illustrate the application of the developed theory, first we consider the subclass of time reversal invariant materials formed by standard isotropic dielectrics characterized by some permittivity $\varepsilon(\omega)$ and some permeability $\mu(\omega)$. Importantly, for such materials the electromagnetic modes can be subdivided into two classes: the TE polarized waves with electric field perpendicular to the plane of propagation ($\mathbf{E} = E_z \hat{\mathbf{z}}$ and $\mathbf{H} = H_x \hat{\mathbf{x}} + H_y \hat{\mathbf{y}}$) and the TM waves with magnetic field perpendicular to the plane of propagation ($\mathbf{H} = H_z \hat{\mathbf{z}}$ and $\mathbf{E} = E_x \hat{\mathbf{x}} + E_y \hat{\mathbf{y}}$) [see Eq. (6)]. The TE and TM waves are completely decoupled, and hence the pseudo-Hamiltonian ($\hat{H}_{cl}$, see Ref. [22]) that characterizes an isotropic dispersive dielectric can be regarded as the sum of two independent Hamiltonians, $\hat{H}_{cl} = \hat{H}_{cl}^{TE} + \hat{H}_{cl}^{TM}$, whereas the whole vector space can be regarded as the direct sum of two independent subspaces (the subspace of TE waves and the subspace of TM waves, respectively). Thus, it is possible to characterize the invariants associated with the TE- waves ($\mathcal{D}^{TE}$) and TM- waves ($\mathcal{D}^{TM}$) separately. The topological invariant of the isotropic dielectric is evidently $\mathcal{D} = \left( \mathcal{D}^{TE} + \mathcal{D}^{TM} \right) \mod 2$. Note that all the invariants ($\mathcal{D}, \mathcal{D}^{TE}, \mathcal{D}^{TM}$) depend on the spectral range $\omega_{min} < \omega < \omega_{max}$, i.e. on the considered subset of photonic bands.

Note that even though the previous discussion deals with isotropic media, it can be readily generalized to uniaxial dielectrics with $\bar{\varepsilon} = \varepsilon_{\parallel}(\omega)(\hat{\mathbf{x}}\hat{\mathbf{x}} + \hat{\mathbf{y}}\hat{\mathbf{y}}) + \varepsilon_{\perp}(\omega)\hat{\mathbf{z}}\hat{\mathbf{z}}$ and $\bar{\mu} = \mu_{\parallel}(\omega)(\hat{\mathbf{x}}\hat{\mathbf{x}} + \hat{\mathbf{y}}\hat{\mathbf{y}}) + \mu_{\perp}(\omega)\hat{\mathbf{z}}\hat{\mathbf{z}}$.

## A.   *Topological invariants for TM-waves*

As a starting point, we consider two distinct materials responses (labeled by the indices 1 and 2) determined by:

$$\varepsilon_1 = 1 - \omega_{pe}^2 / \omega^2, \qquad \mu_1 = 1, \qquad \text{(ENG material)}. \qquad (17a)$$

$$\varepsilon_2 = 1, \qquad \mu_2 = 1 + \omega_{1m}^2 / (\omega_{0m}^2 - \omega^2), \qquad \text{(MNG material)}. \qquad (17b)$$

The material 1 is characterized by a standard Drude dispersion model and has a negative permittivity for $\omega < \omega_{pe}$ and a positive permeability. We will refer to this material as an *epsilon negative* (ENG) material [27]. On the other hand, the permeability of the second material follows a standard Lorentz-type dispersion. In particular, in the frequency range $\omega_{0m} < \omega < \sqrt{\omega_{0m}^2 + \omega_{1m}^2}$ the permeability is negative, while the permittivity is constant and positive. Thus, we refer to the second material as a *mu negative* (MNG) material [27]. It should be noted that in a bandgap an arbitrary isotropic dielectric has necessarily either an ENG or an MNG type response. In general ENG and MNG materials may be designed relying on the metamaterial concept [28].

It is interesting to consider a continuous transformation of the material 1 into the material 2. To this end, we consider a material response dependent on a parameter $\tau$ such that $\mathbf{M}_\tau(\omega) = \mathbf{M}_\infty + (1-\tau)\boldsymbol{\chi}_1(\omega) + \tau\boldsymbol{\chi}_2(\omega)$, with $0 \leq \tau \leq 1$. Here, $\mathbf{M}_\infty = \mathbf{M}(\omega = \infty)$ represents the material matrix of the vacuum, and $\boldsymbol{\chi}_i(\omega) = \mathbf{M}_i(\omega) - \mathbf{M}_\infty$ is the susceptibility associated with the *i*-th material. When the materials 1 and 2 are characterized by parameters consistent with Eq. (17) one has the following explicit dependence of the permittivity and permeability on $\tau$: $\varepsilon_\tau(\omega) = 1 - (1-\tau)\omega_{pe}^2 / \omega^2$ and $\mu_\tau(\omega) = 1 + \tau \omega_{1m}^2 / (\omega_{0m}^2 - \omega^2)$. Thus, the material 1 ($\tau = 0$) is continuously deformed into the material 2 ($\tau = 1$) as the parameter $\tau$ varies in the interval $0 < \tau < 1$.

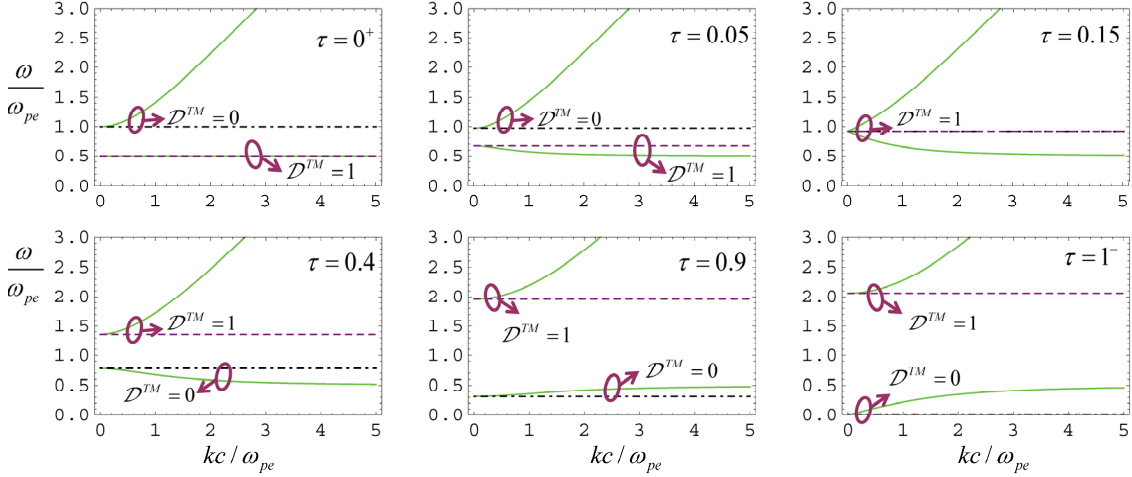

Fig. 2. (Color online) **Topological transition from an ENG material ($\tau = 0$) to an MNG material ($\tau = 1$)** ($\omega_{0m} = 0.5\omega_{pe}$; $\omega_{1m} = 2.0\omega_{pe}$). The purple dashed (black dot-dashed) lines represent the longitudinal modes associated with $\mu = 0$ ($\varepsilon = 0$). The green solid lines are associated with the transverse waves. The insets show the topological invariants for the TM-polarized waves. For $\tau = 0.15$ the bandgap closes and the $\mathbb{Z}_2$ topological index of the upper (lower) bands changes value.

Figure 2 represents the band structure determined by $\mathbf{M}_\tau(\omega)$ for different values of $\tau$, considering that $\omega_{0m} = 0.5\omega_{pe}$ and $\omega_{1m} = 2.0\omega_{pe}$. As is well known, for isotropic materials there are two types of plane waves: transverse waves and longitudinal waves. The transverse waves are doubly degenerate, such that there is a wave associated with TE-polarized modes and a wave associated with TM-polarized modes. On the other hand, the longitudinal waves occur at frequencies where either $\varepsilon = 0$ (longitudinal electric modes, with $\mathbf{E} \sim \hat{\mathbf{k}}$ and $\mathbf{H} = 0$) or $\mu = 0$ (longitudinal magnetic modes, with $\mathbf{H} \sim \hat{\mathbf{k}}$ and $\mathbf{E} = 0$). The longitudinal modes are dispersionless. The longitudinal electric (magnetic) modes are particular cases of TM (TE) polarized waves, respectively.

We calculated the topological invariant for each band subset with TM-polarization ($\mathcal{D}^{TM}$). The value of the topological invariant is indicated in the insets of Fig. 2. Note that for TM-polarization the flat band associated with the longitudinal magnetic modes ($\mu = 0$) must be ignored because it is associated with TE-polarized waves. The calculation of $\mathcal{D}^{TM}$ is done as follows. For the transverse waves we pick the globally defined set of eigenfunctions, $\tilde{\mathbf{f}}_{TM,\mathbf{k}} = \begin{pmatrix} -\dfrac{\mathbf{k}}{\omega_{T,\mathbf{k}} \varepsilon_0 \varepsilon(\omega_{T,\mathbf{k}})} \times \hat{\mathbf{z}} \\ \hat{\mathbf{z}} \end{pmatrix}$ [Eq. (6)], with $\omega = \omega_{T,\mathbf{k}}$ such that $k = \dfrac{\omega}{c}\sqrt{\varepsilon(\omega)\mu(\omega)}$. Typically, these eigenfunctions do not satisfy the Gauge restrictions (12). Indeed, since in the present example the permittivity has no poles for $\omega > 0$, it may be checked that for the branches with $\omega_{T,k=\infty} \neq \infty$ one has $\tilde{\mathbf{f}}_{TM,\mathbf{k}} \approx \begin{pmatrix} -\hat{\mathbf{k}} \times \hat{\mathbf{z}} \\ 0 \end{pmatrix}$ in the limit $k \to \infty$, and hence $\left(\tilde{\mathbf{f}}_{n\mathbf{k}}\right)_{N^-} = -\left(\tilde{\mathbf{f}}_{n\mathbf{k}}\right)_{N^+}$. One can get rid of the undesired minus sign with a smooth Gauge transformation, $\mathbf{f}_{TM,\mathbf{k}} = \tilde{\mathbf{f}}_{TM,\mathbf{k}} e^{i\theta_{\mathbf{k}}}$, with $\theta_{\mathbf{k}}$ such that $\left(\theta_{\mathbf{k}}\right)_{N^+} - \left(\theta_{\mathbf{k}}\right)_{N^-} = \pi$ [‡], so that $\left(\mathbf{f}_{n\mathbf{k}}\right)_{N^-} = \left(\mathbf{f}_{n\mathbf{k}}\right)_{N^+}$. Because $\tilde{\mathbf{f}}_{TM,\mathbf{k}}$ is real valued the corresponding Berry potential vanishes. Thus, $\mathcal{A}_{\mathbf{k}} = -\nabla_{\mathbf{k}} \theta_{\mathbf{k}}$ and from Eq. (14) one sees that each TM transverse branch with $\omega_{T,k=\infty} \neq \infty$ gives a contribution (calculated over the half-circle centered at the $N$ point) of +1 to the invariant. Using similar arguments one can check that the contribution of the longitudinal-electric branch ($\tilde{\mathbf{f}}_{LE,\mathbf{k}} \sim \begin{pmatrix} \hat{\mathbf{k}} \\ 0 \end{pmatrix}$) and of the transverse branch with $\omega_{T,k=\infty} = \infty$ at the $N$ point is also +1. On the other hand, it may be verified that the individual contributions from the transverse and longitudinal branches at the $S$ pole are in general nonzero. However, the total contribution from all branches at the $S$ pole vanishes. Indeed, as discussed in Sect. II.D the south-pole can only yield a nontrivial contribution to the $\mathbb{Z}_2$ invariant for bands with $\omega_{n,k=0} = 0$ (in the examples of Fig. 2 all the bands have $\omega_{n,k=0} > 0$). Notably, the previous analysis indicates that for isotropic dielectrics (which have a response that stays invariant under the inversion transformation $\mathbf{r} \to -\mathbf{r}$) the value of the topological invariant for each eigenmode branch is intrinsically related to the parity (odd or even) of the electric field at the time-reversal invariant points of the wave vector space. A similar property applies to electronic topological insulators with the inversion symmetry [26].

---

[‡] For example, one may pick $\theta_{\mathbf{k}} = \dfrac{\pi}{2}(\cos(\varphi)+1)e^{-1/k}$, where $(k,\varphi)$ determines a system of polar coordinates in the **k**-plane, so that $\left(\theta_{\mathbf{k}}\right)_{\varphi=0^+,k=\infty} - \left(\theta_{\mathbf{k}}\right)_{\varphi=\pi,k=\infty} = \pi$. Note that this $\theta_{\mathbf{k}}$ does not change the behavior of the eigenfunctions near the $S$ pole ($k=0$).

As seen in Fig. 2, the high-frequency photonic bands have a topological index $\mathcal{D}^{TM}$ different from the low-frequency bands. Importantly, as $\tau$ varies from $0^+$ (ENG-type material response) to $1^-$ (MNG-type material response) the topological invariant of the high-frequency and low-frequency bands is interchanged. The topological transition takes place precisely when $\tau = 0.15$, i.e. when the bandgap closes and $\varepsilon_\tau = \mu_\tau = 0$. Note that for $\tau = 0.15$ the bands with $\omega > \sqrt{1-\tau}\omega_{pe} = 0.92\omega_{pe}$ have $\varepsilon, \mu > 0$ (double positive – DPS – material response), whereas the bands with $\omega < 0.92\omega_{pe}$ have $\varepsilon, \mu < 0$ (double negative – DNG – material response). This result demonstrates that if one considers *only*

TM-polarized waves then the ENG material (material with $\tau = 0^+$) is topologically distinct from the MNG material (material with $\tau = 1^-$). Note that the two materials share a common bandgap, $\omega_{0m} < \omega < \omega_{pe}$, and that the topological numbers associated with the eigenmode branches in any of the two intervals $\omega_{gap} < \omega < +\infty$ or $0^+ < \omega < \omega_{gap}$ are distinct in the two materials, i.e. $\Delta \mathcal{D}^{TM} = 1$ where $\Delta \mathcal{D}^{TM} = \left( \mathcal{D}^{TM}_{MNG} - \mathcal{D}^{TM}_{ENG} \right) \mod 2$. Here, $\omega_{gap}$ is any frequency in the common bandgap.

It is relevant to note at this point that the dimension of the Hilbert space wherein the pseudo-Hamiltonian of a given continuous medium is defined depends on the number of poles of the material response [22]. Hence, if one wishes to compare two generic materials (characterized by the material matrices $\mathbf{M}_1$ and $\mathbf{M}_2$), and test if they are topologically equivalent or not it is necessary to choose an underlying Hilbert space that is common to the two materials. This can be done using the combined material matrix $\mathbf{M}_\tau(\omega)$, which effectively merges the poles of the two material responses. Hence, we say that two materials $\mathbf{M}_1$ and $\mathbf{M}_2$ are topologically equivalent (inequivalent) in some spectral range if the topological numbers of $\mathbf{M}_{\tau=0^+} \approx \mathbf{M}_1$ and $\mathbf{M}_{\tau=1^-} \approx \mathbf{M}_2$ are the same (different). The finding that ENG and MNG materials are topologically distinct ($\Delta \mathcal{D}^{TM} = 1$) is consistent with this definition.

## B.  *Topological invariants for TE-waves and for the bulk dielectric*

The topological numbers for TE-polarized waves can be calculated using a procedure analogous to that outlined in the previous subsection. For transverse waves we start with the globally defined Gauge $\tilde{\mathbf{f}}_{TE,\mathbf{k}} = \begin{pmatrix} \hat{\mathbf{z}} \\ \dfrac{\mathbf{k}}{\omega_{T,\mathbf{k}}\mu_0 \mu(\omega_{T,\mathbf{k}})} \times \hat{\mathbf{z}} \end{pmatrix}$ [Eq. (6)]. It can be checked that this Gauge gives $\left(\tilde{\mathbf{f}}_{TE,\mathbf{k}}\right)_{N^-} = \left(\mathcal{U} \cdot \tilde{\mathbf{f}}_{TE,\mathbf{k}}\right)_{N^+}$ for the high-frequency transverse band with $\omega_{T,k=\infty} = \infty$, and $\left(\tilde{\mathbf{f}}_{TE,\mathbf{k}}\right)_{N^-} = \left(\tilde{\mathbf{f}}_{TE,\mathbf{k}}\right)_{N^+}$ for the low-frequency transverse band with $\omega_{T,k=\infty}$ finite (note that for this band the permeability has a pole). Thus, the behavior of $\tilde{\mathbf{f}}_{TE,\mathbf{k}}$ near the $N$ point is compatible with the Gauge restrictions (12). Because the Berry potential associated with $\tilde{\mathbf{f}}_{TE,\mathbf{k}}$ vanishes, we conclude that the contribution to $\mathcal{D}^{TE}$ from the transverse bands at the $N$ point vanishes. On the other hand, it may be checked that the contribution from the longitudinal magnetic-branch ($\tilde{\mathbf{f}}_{LM,\mathbf{k}} \sim \begin{pmatrix} 0 \\ \hat{\mathbf{k}} \end{pmatrix}$) to $\mathcal{D}^{TE}$ at the $N$ point is +1. Again, the $S$ point gives no contributions to the invariant. Using these results it can be verified that $\mathcal{D}^{TE} = \mathcal{D}^{TM}$ in all the cases of Fig. 2. Thus, it follows that also for TE-polarized waves the ENG material ($\tau = 0^+$) is topologically distinct from the MNG material ($\tau = 1^-$).

An immediate consequence of this finding is that the topological number for the bulk dielectric (including both TE- and TM- polarized waves) is $\mathcal{D} = \left(\mathcal{D}^{TE} + \mathcal{D}^{TM}\right) \mod 2 = 0$. In other words, a bulk isotropic dielectric is always (in any frequency interval of interest) topologically trivial. Thus, the Stokes theorem can be applied with no obstructions to the $EBZ$. However, if one restricts himself to either TE- or TM-polarized waves, then in general there are obstructions to the application of the Stokes theorem, and an isotropic dielectric may be topologically nontrivial in the subspace of Hamiltonians associated with a specific wave polarization.

Even though bulk isotropic dielectrics are topologically trivial (when there are no polarization restrictions), it is relevant to note that bulk anisotropic uniaxial dielectrics can be topologically nontrivial. Indeed, for uniaxial media $\mathcal{D}^{TE}$ ($\mathcal{D}^{TM}$) only depends on the permittivity and permeability components $\varepsilon_\perp, \mu_\parallel$ ($\varepsilon_\parallel, \mu_\perp$). Thus, in general the band structures of the TE and TM polarizations can be totally independent from each other, and hence one can obviously have $\mathcal{D} = \left(\mathcal{D}^{TE} + \mathcal{D}^{TM}\right) \bmod 2$ different from zero for some eigenmode branches.

## C.  *Topological edge states*

It is natural to wonder if similar to electronics [4, 5] having two different $\mathbb{Z}_2$ topological numbers in two bulk materials ($\Delta\mathcal{D}=1$) may imply the existence of topologically protected "helical" edge states. Specifically, in electronic systems it is known that if the interface (let us say along the *x*-direction) between two $\mathcal{T}$-invariant electronic materials with a common bandgap supports an odd number of edge states in half of the Brillouin zone ($0 < k_x < \pi/a$, being *a* the period along the *x*-direction and $k_x$ the propagation constant of the edge modes), then the Kramers theorem ensures the topological protection of these states [4]. Indeed, any continuous deformation of the two bulk materials that preserves the time-reversal invariance and does not close the common bandgap can only change the number of edge states modulo 2 [4].

Unfortunately, despite some parallelisms discussed in Appendix B, in photonic systems there is no precise analogue of the Kramers theorem. This suggests that the dispersion of the edge states in photonic systems cannot be protected in the same robust topological manner as in electronics. Can we however make sense of the topological inequivalence of two materials ($\Delta \mathcal{D}=1$) in the context of the edge states propagation?

Suppose that two topologically inequivalent photonic materials ($\mathbf{M}_1$ and $\mathbf{M}_2$) are put side by side. One can regard the interface as a very thin layer wherein the pseudo Hamiltonian of the material $\mathbf{M}_1$ is continuously deformed into the material matrix $\mathbf{M}_2$ [5]. Because the materials are topologically inequivalent the bandgap must close and eventually reopen somewhere in the thin interfacial layer. Hence, this suggests that the two topologically inequivalent materials must support edge states *somewhere* in the common bandgap. Crucially, different from electronic systems, the edge states dispersion *is not* required to span the entire bandgap. In electronic systems the Kramers theorem guarantees that if the number of edge states modulo 2 is nonzero then the edge states dispersion spans the entire gap [4]. This happens because the Kramers theorem (when applied to the heterostructure) forbids the edge modes dispersion from opening a gap at the $\mathcal{T}$-invariant points $k_x = 0, k_x = \pi/a$ [4]. However, as previously discussed, in photonic systems there is no exact analogue of the Kramers theorem.

In summary, the previous heuristic discussion suggests that the bulk-edge correspondence for photonic topologically inequivalent $\mathcal{T}$-invariant materials only allows establishing the emergence of topological edge states in the common bandgap. The topological edge states are not required to span the entire bandgap.

To test these ideas we studied the dispersion of the edge modes supported by an interface between an ENG ($\varepsilon_1, \mu_1$) and an MNG material ($\varepsilon_2, \mu_2$). As seen in the previous subsections, these materials are topologically different when the wave propagation is restricted to be either TE or TM polarized. As is well known, the dispersion of the guided modes for TE (TM) polarization is determined by the characteristic equation $\frac{\gamma_1}{\mu_1} + \frac{\gamma_2}{\mu_2} = 0$ ($\frac{\gamma_1}{\varepsilon_1} + \frac{\gamma_2}{\varepsilon_2} = 0$) where $\gamma_i = \sqrt{k_x^2 - \omega^2 \varepsilon_i(\omega)\mu_i(\omega)/c^2}$, $i=1,2$, and $k_x$ is the propagation constant of the guided mode along the interface (directed along the $x$-direction).

Let us focus on the TM polarization case, and prove that, indeed, there is always a branch of edge modes lying within the common bandgap of the two topologically different materials. Specifically, next we prove that there is always an edge mode with $k_x = 0$ (obviously, this edge mode is analytically continued to form a branch of edge modes). Suppose that the common bandgap is defined by $\omega_L < \omega < \omega_U$ where $\omega_L, \omega_U$ determine the bandgap edges. Then, for $k_x = 0$ the TM dispersion equation for frequencies within the common bandgap reduces to $f(\omega) = 0$ with $f \equiv -\sqrt{\frac{\mu_1}{|\varepsilon_1|}} + \sqrt{\frac{|\mu_2|}{\varepsilon_2}}$.

We used the fact that in the common bandgap $\varepsilon_1 < 0, \mu_1 > 0$ (ENG material) and $\varepsilon_2 > 0, \mu_2 < 0$ (MNG material). Clearly, $\omega_L, \omega_U$ must satisfy one the following four conditions: (i) $\varepsilon_1(\omega_U) = 0$ and $\varepsilon_1(\omega_L) = \infty$. (ii) $\mu_2(\omega_U) = 0$ and $\mu_2(\omega_L) = \infty$. (iii) $\varepsilon_1(\omega_U) = 0$ and $\mu_2(\omega_L) = \infty$. (iv) $\mu_2(\omega_U) = 0$ and $\varepsilon_1(\omega_L) = \infty$. Notably, it can be checked that for all the four cases one has $f(\omega_L) f(\omega_U) < 0$. For example in the case (i) $f(\omega_L) = \sqrt{\frac{|\mu_2|}{\varepsilon_2}}\bigg|_{\omega_L} > 0$ and $f(\omega_U) = -\infty$. The property $f(\omega_L) f(\omega_U) < 0$ demonstrates that there is, indeed, some frequency in the gap for which $f(\omega) = 0$, and this proves the desired result. Analogously, one can verify that for TE-polarization there is always an edge mode branch in the common bandgap. Thus, the bulk edge correspondence discussed in the beginning of this subsection really applies to the case of an interface between ENG and MNG materials.

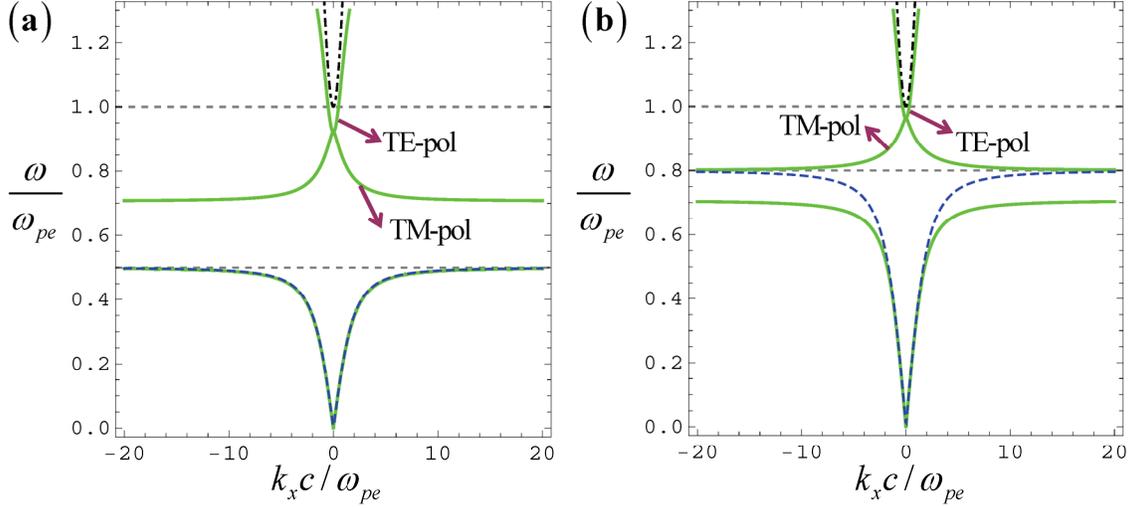

Fig. 3. (Color online) **Dispersion of the edge modes at an interface between an ENG material and an MNG material.** The dashed horizontal gray lines delimit the common bandgap of the two bulk materials. The solid green lines represent the edge modes dispersion. The TE and TM branches within the bandgap are identified by the arrows and meet at the $k_x = 0$ point. The dashed blue (dot-dashed black) lines represent the dispersion of the transverse modes of the bulk MNG (ENG) materials with material dispersion as in Eq. (17) and (a) $\omega_{0m} = 0.5\omega_{pe}$; $\omega_{1m} = 2.0\omega_{pe}$. (b) $\omega_{0m} = 0.8\omega_{pe}$; $\omega_{1m} = 2.0\omega_{pe}$.

To illustrate the discussion, we show in Fig. 3a the dispersion of the edge modes for both TE and TM polarizations and for the same ENG and MNG materials as in Fig. 2. Figure 3b shows a similar plot for the case wherein the resonant frequency of the MNG material is increased to $\omega_{0m} = 0.8\omega_{pe}$. As seen, in both examples there are always topologically protected edge modes in the common gap (delimited by the gray horizontal lines) for both polarizations. Importantly, the dispersion of the individual TE (TM) edge modes does not fully span the entire gap. Note that the TE and TM branches meet at the point wherein $k_x = 0$, and that there are also guided modes outside the common bandgap.

At this point, it is relevant to mention that the surface plasmons polaritons (SPPs) supported by metal-air interfaces are not part of our theory. Indeed, air is a transparent material while the topological index can only defined for "light insulators" i.e., for materials that do not support light states in some bandgap (opaque materials).

The topological edge modes supported by an ENG-MNG interface have a "helical" nature due to the universal transverse spin-momentum locking characteristic of evanescent waves, which has been linked to a quantum spin Hall effect for light and a spin-orbit interaction in optics [29-30]. Several recent studies demonstrated highly robust spin-assisted unidirectional excitation of edge modes at metallic interfaces [31-33]. To demonstrate the helical nature of the topological ENG-MNG edge states, we consider a 2D kite-shaped ENG object embedded in an MNG material (Fig. 4). The topological TE edge modes are excited by an in-plane magnetic current density of the form $\mathbf{j}_m = -i\omega \mathbf{p}_m \delta(x-x_0)\delta(y-y_0)$, where $(x_0, y_0)$ are the coordinates of the point source and $\mathbf{p}_m \sim \hat{\mathbf{x}} \pm i\hat{\mathbf{y}}$ is the magnetic dipole moment per unit of length. This source favors the excitation of edge modes for which the in-plane magnetic field rotates with a specific helicity determined by the ± sign. Thus, because of the universal transverse spin-momentum locking [29-30], it is possible to excite only the modes that match the helicity of the source and have a unidirectional light flow. This is illustrated in Fig. 4 and also in field animations available in the supplementary materials [34] for two cases: (*i*) the structure is lossless. (*ii*) a lossy circular object is centered at $(x_{obj}, y_{obj}) = (2.51\lambda_0, 0)$ in the vicinity of the kite-shaped object. In both cases it is evident that the source excites mainly an edge mode that flows in the counterclockwise direction around the kite-shaped object (see in particular the field animations in Ref [34]). In the case *(ii)*, the launched wave is strongly absorbed by the lossy object. Note that because the two bulk materials do not support light states the radiation is strongly confined to the interface. The numerical simulations were done using the Nyström method [35], which is an integral equation approach related to the Galerkin method.

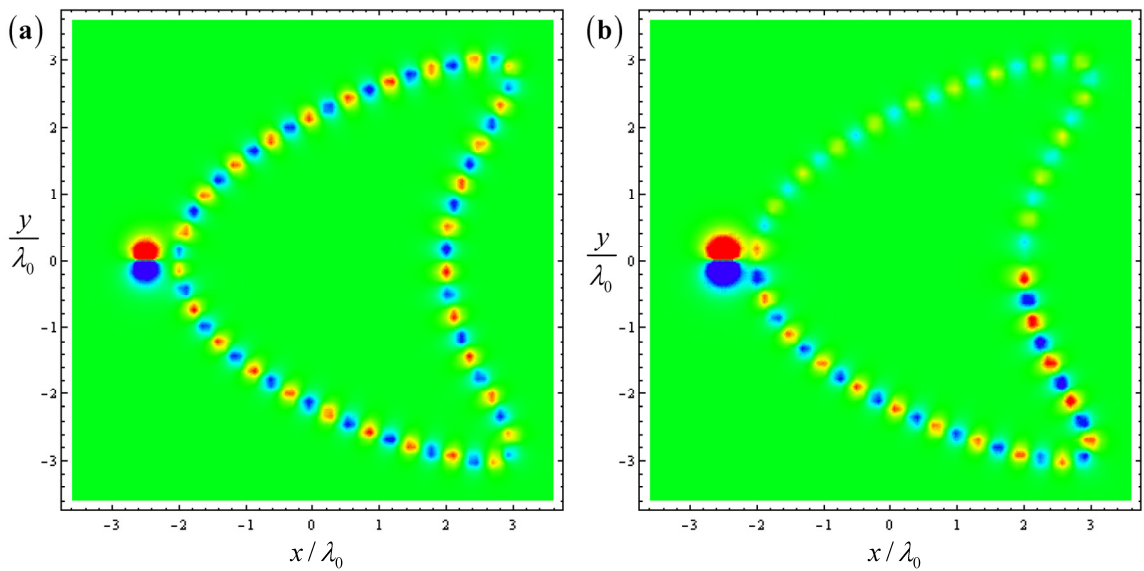

**Fig. 4**. (Color online) **Spin assisted unidirectional excitation of edge modes at an MNG-ENG interface.** The background material is an MNG material with $\omega_{0m} = 0.5\omega_{pe}$; $\omega_{1m} = 1.2\omega_{pe}$ and the kite-shaped object is made of a ENG material with plasma frequency $\omega_{pe}$. The frequency of operation is $0.95\omega_{pe}$ and $\lambda_0$ is the corresponding vacuum wavelength. The light source is positioned at $(-2.5\lambda_0, 0)$ and excites a TE-polarized edge mode that propagates in the counterclockwise direction. (a) time snapshot of $E_z$ for a lossless structure. The bluish (reddish) colors represent positive (negative) values of the field. The green color represents a vanishing field. (b) time snapshot of $E_z$ when a lossy circular object with $R = 0.5\lambda_0$ is centered at $(x_{obj}, y_{obj}) = (2.51\lambda_0, 0)$. The parameters of the lossy object are the same as those of the MNG background except that the imaginary parts of the material parameters are $\varepsilon'' = \mu'' = 0.3$.

To conclude this section, it is relevant to discuss if similar to electronics the edge states are topologically protected against back-scattering. Unfortunately, the answer to this question appears to be negative. The reason is simple to explain, and again it is deeply rooted in the fact that $\mathcal{T}^2=-\mathbf{1}$ for fermionic systems, whereas $\mathcal{T}^2=\mathbf{1}$ for bosonic systems. Indeed, it can be shown that this property implies that the scattering matrix satisfies $\mathbf{S}=-\mathbf{S}^T$ in the electronic case [36], whereas in the photonic case one has instead $\mathbf{S}=+\mathbf{S}^T$ [37]. The different leading sign in the two cases has tremendous consequences. In fact, the property $\mathbf{S}=-\mathbf{S}^T$ guarantees that when there are an odd number of edge states (and hence – because of the Kramers theorem –the edge states dispersion is gapless) there is at least one conducting state that is transmitted with no reflections [36]. This is valid for any type of perturbation that does not break the time-reversal symmetry. Unfortunately, this extraordinary result has no correspondence in photonics because in this case the time-reversal symmetry gives $\mathbf{S}=\mathbf{S}^T$, and hence independent if the number of propagating edge states is odd or even it is impossible to guarantee propagation immune to back-scattering.

## IV. Topologically Nontrivial Media with Magnetoelectric Coupling

It was seen in Sect. III that isotropic dielectrics are always topologically trivial when the wave polarization is arbitrary, i.e. when the polarization is not restricted to either the TE or TM cases. As mentioned in the end of subsection III.B, the simplest example of a topologically nontrivial material is the case of a uniaxial dielectric.

A magneto-electric (chiral-type) response also gives the opportunity to have nontrivial topological numbers (without any polarization restrictions). To illustrate this, next we study the topological invariants of the so-called (pseudo-chiral) $\Omega$-material introduced by Saadoun and Engheta [38, 39]. Our motivation is that the photonic topological insulators introduced in Ref. [6] also rely on a $\Omega$-type reciprocal magneto-electric coupling. Specifically, we consider an $\Omega$-material with the general $\mathcal{T}$-invariant material matrix:

$$\mathbf{M} = \begin{pmatrix} \varepsilon_0 \varepsilon \mathbf{1}_{3\times3} & -i\Omega \frac{1}{c} \hat{\mathbf{z}} \times \mathbf{1}_{3\times3} \\ -i\Omega \frac{1}{c} \hat{\mathbf{z}} \times \mathbf{1}_{3\times3} & \mu_0 \mu \mathbf{1}_{3\times3} \end{pmatrix}, \tag{18a}$$

such that the permittivity, the permeability, and the $\Omega$-parameter satisfy:

$$\varepsilon = 1 + \frac{\omega_e^2}{\omega_0^2 - \omega^2}, \qquad \mu = 1 + \frac{\omega_m^2}{\omega_0^2 - \omega^2}, \qquad \Omega = \frac{\omega \omega_\Omega}{\omega_0^2 - \omega^2}. \tag{18b}$$

Here, $\omega_0$ is the resonance frequency, and $\omega_e, \omega_m, \omega_\Omega$ determine the strengths of the electric, magnetic, crossed electromagnetic resonances, respectively. In order that the restrictions enunciated in Sect. II.A are satisfied it is necessary that $|\omega_\Omega| < \omega_e \omega_m / \omega_0$. The plane waves in the $\Omega$-material are *not* transverse waves. Yet, it can be shown that the $\Omega$-material supports two degenerate transverse-type waves that satisfy the dispersion equation:

$$k^2 = \frac{\omega^2}{c^2} \left( \varepsilon \mu - \Omega^2 \right). \tag{19}$$

It is possible to choose a basis for these eigenmodes formed by TE and TM-type waves:

$$\mathbf{f}_{TM,\mathbf{k}} \sim \begin{pmatrix} \mathbf{k} \times \hat{\mathbf{z}} \\ \frac{1}{\mu_0 \mu} \left( \frac{-k^2}{\omega} \hat{\mathbf{z}} + i\Omega \frac{1}{c} \mathbf{k} \right) \end{pmatrix}, \qquad \mathbf{f}_{TE,\mathbf{k}} \sim \begin{pmatrix} \frac{-1}{\varepsilon_0 \varepsilon} \left( \frac{-k^2}{\omega} \hat{\mathbf{z}} - i\Omega \frac{1}{c} \mathbf{k} \right) \\ \mathbf{k} \times \hat{\mathbf{z}} \end{pmatrix}. \tag{20a}$$

Importantly, these eigenmodes are *not* pure TE and TM waves. For example, the magnetic field associated with $\mathbf{f}_{TM,\mathbf{k}}$ is not perpendicular to the plane of propagation when $\Omega \neq 0$. In addition, the $\Omega$-material supports non-degenerate longitudinal-type dispersionless waves that occur at frequencies for which either $\varepsilon = 0$ (longitudinal-type electric –LE– mode with $\mathbf{H} = 0$) or $\mu = 0$ (longitudinal-type magnetic –LM– mode with $\mathbf{E} = 0$). The longitudinal-type modes are associated with the eigenfunctions:

$$\mathbf{f}_{LE,\mathbf{k}} \sim \begin{pmatrix} \dfrac{\mathbf{k}}{\omega} + i\Omega \dfrac{1}{c}\hat{\mathbf{z}} \\ \mathbf{0} \end{pmatrix}, \qquad \mathbf{f}_{LM,\mathbf{k}} \sim \begin{pmatrix} \mathbf{0} \\ \dfrac{\mathbf{k}}{\omega} - i\Omega \dfrac{1}{c}\hat{\mathbf{z}} \end{pmatrix}. \qquad (20b)$$

The modes LE and LM are not purely longitudinal because for $\Omega \neq 0$ they also have a $z$ component.

Using these results we studied the topological transition from an $\Omega$-material with material matrix as in Eq. (18) (material matrix $\mathbf{M}_1$) to a standard ENG material with $\varepsilon_2 = 1 - \omega_{pe}^2/\omega^2$ and $\mu_2 = 1$ (material matrix $\mathbf{M}_2$). The material matrix $\mathbf{M}_\tau(\omega)$ is used to link the two material responses $\mathbf{M}_1$ and $\mathbf{M}_2$, being $\mathbf{M}_\tau$ defined in the same way as in Sect. III. It can be checked that $\mathbf{M}_\tau$ is always of the form (18a), and hence it determines the dispersion of an $\Omega$-material that is some interpolation of $\mathbf{M}_1$ and $\mathbf{M}_2$. In Fig. 5 we depict the evolution of the band structure as the material response changes continuously from $\mathbf{M}_{\tau=0^+} = \mathbf{M}_1$ ($\Omega$-material) to $\mathbf{M}_{\tau=1^-} = \mathbf{M}_2$ (ENG material). The green solid lines represent the transverse-type modes and are doubly degenerate. The dashed and dot-dashed flat lines represent the non-degenerate longitudinal-type modes (LM and LE, respectively).

The $\mathbb{Z}_2$ topological index $\mathcal{D}$ associated with each band subset can be calculated as explained next. First, we note that for the same reasons as in Sect. III the $\mathcal{D}$ number is determined uniquely by the eigenfunctions behavior near the north-pole ($k = \infty$), i.e. the individual contributions from the south-pole always add up to zero. In the north-pole there are three relevant band types. The first type corresponds to the two high-frequency bands for which $\omega_{n,k=\infty} \to \infty$. Because in the $\omega \to \infty$ limit the $\Omega$-parameter vanishes, $\mathbf{f}_{TM,\mathbf{k}}$ and $\mathbf{f}_{TE,\mathbf{k}}$ in Eq. (20a) have the same asymptotic behavior as the branches $\mathbf{f}_{TM,\mathbf{k}}$ and $\mathbf{f}_{TE,\mathbf{k}}$ in Eq. (6) for an isotropic dielectric. Hence, from Sects. III.A and III.B it follows that the invariant for the two degenerate high frequency bands is $\mathcal{D} = 0+1 = 1$. The second relevant band-type consists of the non-degenerate LE and LM modes. It is seen from Eq. (20b) that in the $k \to \infty$ limit the LE and LM modes become purely longitudinal. Hence, from Sect. III (see also Appendix C) the $\mathbb{Z}_2$ index for each longitudinal-type branch is $\mathcal{D} = 1$. Finally, the last set of relevant bands are the bands associated with the resonance of the $\Omega$-coupling. For example, in the panel $\tau = 0^+$ in Fig. 5 these bands (formed by two different green lines, being each green line doubly degenerate) are such that $\omega_{n,k=\infty} = \omega_0 = 0.5\omega_{pe}$. The results of Appendix C [subsection III] show that these four transverse-type modes give a vanishing contribution to $\mathcal{D}$ because the number $N$ of eigenmode branches with $\omega_{n,\mathbf{k}} = \omega_{n,k=\infty} + 0^+$ in the limit $k \to \infty$ is even (N=2).

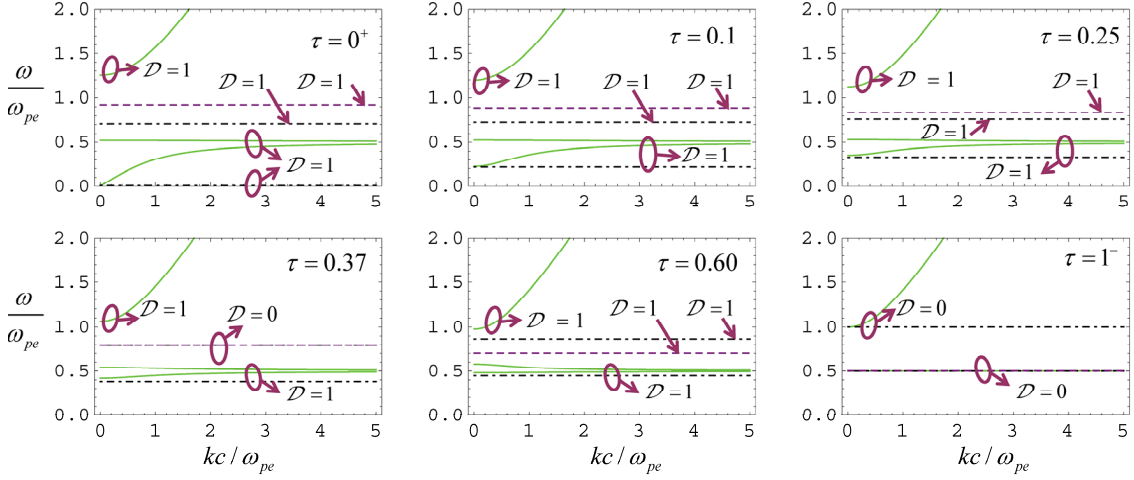

Fig. 5. (Color online) **Topological transition from an Ω-medium ($\tau = 0$) to an ENG material ($\tau = 1$)** ($\omega_0 = 0.5\omega_{pe}$, $\omega_e = \omega_0$, $\omega_m = 1.55\omega_0$, $\omega_\Omega = 0.9\omega_e\omega_m/\omega_0$). The purple dashed (black dot-dashed) lines represent the dispersion of the longitudinal-type modes associated with $\mu = 0$ ($\varepsilon = 0$). The green solid lines represent the doubly degenerate transverse-type modes.

The band diagrams of Fig. 5 reveal that the transition from the $\Omega$-material to an ENG material involves merging some frequency branches and adding up the corresponding topological invariants. Clearly, the $\Omega$-material is topologically nontrivial in some spectral ranges. By comparing the band diagrams of the panels $\tau = 0^+$ and $\tau = 1^-$, we see that the $\Omega$-material and the ENG material have three common bandgaps *(i)* $\omega_{LM} < \omega < \omega_{pe}$ *(ii)* $\omega_{LE} < \omega < \omega_{LM}$ and *(iii)* $\omega_0 < \omega < \omega_{LE}$, where $\omega_{LM} = 0.92\omega_{pe}$ and $\omega_{LE} = 0.71\omega_{pe}$ are the frequencies associated with the high-frequency longitudinal-type LM and LE modes and $\omega_0 = 0.5\omega_{pe}$. For a given common bandgap, let $\mathcal{D}_\Omega$ and $\mathcal{D}_{ENG}$ denote the invariants of the photonic bands with dispersion below the gap for the $\Omega$- and ENG- materials, respectively. We put $\Delta\mathcal{D} = (\mathcal{D}_\Omega - \mathcal{D}_{ENG}) \bmod 2$. Then, it is easy to check that $\Delta\mathcal{D} = 1$ for the common bandgaps *(i)* and *(iii)*, whereas $\Delta\mathcal{D} = 0$ for the bandgap *(ii)*. Therefore, the bulk-edge correspondence discussed in Sect. III.C predicts that an $\Omega$-

ENG interface supports topologically protected edge states in the two bandgaps *(i)* and *(iii)*, i.e. when the two materials have a different $\mathbb{Z}_2$ topological index.

To confirm this prediction we computed the edge states for an interface of the $\Omega$- and ENG- materials along the *x*-direction. The dispersion of the edge states is of the form (the details of the derivation are omitted for conciseness):

$$\left(\gamma_\Omega + \frac{\gamma_{iso}}{\varepsilon_{iso}}\frac{n_\Omega^2}{\mu}\right)\left(\gamma_\Omega + \frac{\gamma_{iso}}{\mu_{iso}}\frac{n_\Omega^2}{\varepsilon}\right) - \frac{\Omega^2 k_x^2}{\varepsilon\mu} = 0, \qquad (21)$$

where $\varepsilon, \mu, \Omega$ are the parameters of the $\Omega$-medium and $\varepsilon_{iso} = \varepsilon_2$, $\mu_{iso} = \mu_2$ stand for the parameters of the ENG material. In the above, $\gamma_{iso} = \sqrt{k_x^2 - \varepsilon_{iso}\mu_{iso}\omega^2/c^2}$, $\gamma_\Omega = \sqrt{k_x^2 - n_\Omega^2\omega^2/c^2}$, and $n_\Omega^2 = \varepsilon\mu - \Omega^2$. The edge modes cannot be classified as either TE or TM, and in general are hybrid modes.

Figure 6a depicts the calculated dispersion for the edge modes (green lines) when the $\Omega$- and ENG- materials of Fig. 5 are put side-by-side. As seen, consistent with the bulk-edge correspondence, the interface supports topological edge states in the two bandgaps *(i)* and *(iii)* (the two shaded yellow regions of Fig. 6a). Actually, in this example there are also edge states in the bandgap *(ii)*. However, the topological protection only applies to the bandgaps *(i)* and *(iii)*. This is made clear in Fig. 6b, which depicts the edge modes dispersion at an interface between the same ENG material and a different $\Omega$- material. In this configuration there are also three common bandgaps, but again the topological protection only applies to two of them (shaded yellow regions in Fig. 6b). This second example further demonstrates the validity of the bulk-edge correspondence and confirms that the topological protection applies only to selected common bandgaps with $\Delta\mathcal{D}=1$.

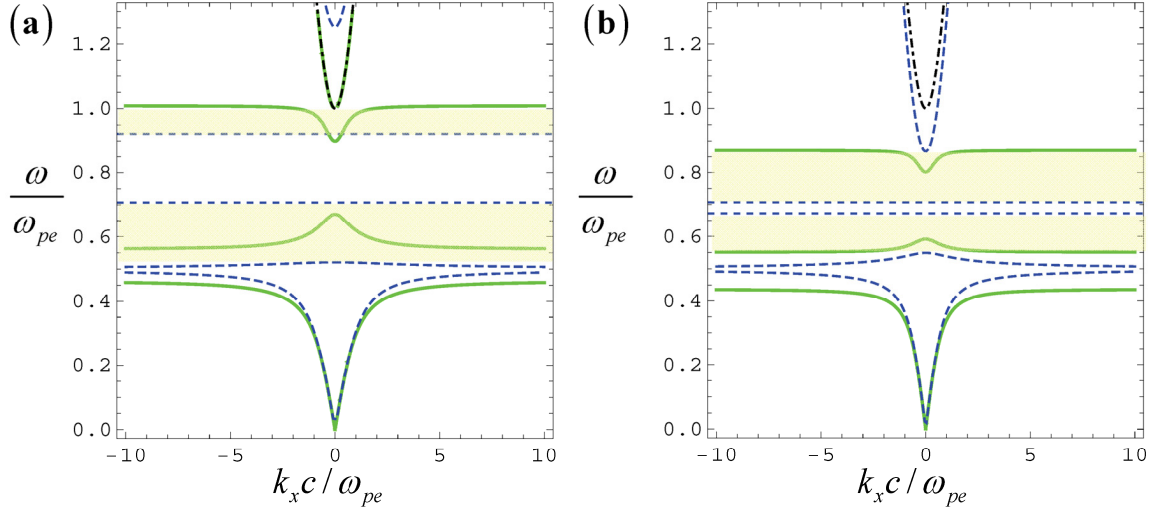

Fig. 6. (Color online) **Dispersion of the edge modes at an interface between an ENG material and an Ω-medium.** The shaded yellow regions delimit the common bandgaps wherein the topological edge modes are predicted to occur. The solid green lines represent the edge modes dispersion. The dashed blue (dot-dashed black) lines represent the dispersion of the transverse-type modes of the bulk Ω-material (ENG material). The two dashed horizontal lines represent the dispersion of longitudinal-type modes of the Ω-medium. (a) $\omega_0 = 0.5\omega_{pe}$, $\omega_e = \omega_0$, $\omega_m = 1.55\omega_0$, $\omega_\Omega = 0.9\omega_e\omega_m/\omega_0$. (b) $\omega_0 = 0.5\omega_{pe}$, $\omega_e = \omega_0$, $\omega_m = 0.9\omega_0$, $\omega_\Omega = 0.7\omega_e\omega_m/\omega_0$.

## V. Discussion and Conclusion

We introduced a $\mathbb{Z}_2$ topological index for continuous photonic systems with the time reversal symmetry. A nontrivial $\mathbb{Z}_2$ index implies an obstruction to the application of the Stokes theorem to the effective Brillouin zone. The $\mathbb{Z}_2$ topological number generally depends on the considered plane of propagation, particularly if the bandgaps in different planes do not overlap. Such a situation may occur in general bi-anisotropic media. Our theory shows that the nontrivial contributions to the $\mathbb{Z}_2$ index are related to either the non-smooth nature of the pseudo-Hamiltonian at the north-pole or to the entanglement of positive and negative frequency branches. Our $\mathbb{Z}_2$ topological index generalizes to photonics the notion of a topological insulator in periodic electronic systems. The formulas of the $\mathbb{Z}_2$ index in electronics and photonics differ by a factor of two because of the different spins of electrons and photons.

In principle, it is possible to generalize the ideas of this article also to dielectric photonic crystals, i.e. to systems with an intrinsic periodicity. It should be noted that for photonic crystals formed by *nondispersive* dielectrics the pseudo-Hamiltonian is expected to be smooth in the entire Brillouin zone. If that is the case, it appears that a nonzero $\mathbb{Z}_2$ topological index can only arise due to a nontrivial entanglement of the positive and negative frequency branches. The characterization of the $\mathbb{Z}_2$ index of photonic crystals requires thus further studies.

It was demonstrated that isotropic dielectrics are always topologically trivial. Yet, when the wave propagation is restricted to either TE- or TM- polarization an isotropic dielectric may have topologically nontrivial features. As a consequence of this, an interface between an ENG material and an MNG material always supports topologically protected edge states for both TE and TM polarizations. Nevertheless, different from electronics and due to the absence of Kramers partners, the dispersion of the edge states does not have to span the entire gap. The bulk-edge correspondence only ensures that for a continuous deformation of the materials that does not close the common gap there are always edge states *somewhere* in the gap. It was shown that the bulk-edge correspondence is exact for isotropic dielectrics, and it is an open question if this property is universal or not. Moreover, it was numerically demonstrated that the "helical"-nature of the topological edge states enables the unidirectional emission of light. However, different from electronics the topological edge states are not protected against back-scattering because the time reversal symmetry imposes that the scattering matrix satisfies $\mathbf{S} = \mathbf{S}^T$ for light waves, whereas for electron waves one has instead $\mathbf{S} = -\mathbf{S}^T$. Finally, it was demonstrated that continuous media with magnetoelectric coupling can also be topologically nontrivial. We hope that the developed ideas can contribute to a more complete understanding of the use of topological methods in photonics and to the design of novel ultra-compact photonic platforms for light localization and guiding relying on single interface edge modes.

**Acknowledgements:** This work is supported in part by Fundação para a Ciência e a Tecnologia grants number PTDC/EEI-TEL/2764/2012 and PTDC/EEI-TEL/4543/2014.

## *Appendix A: Proof that the eigenspaces are continuous at the north-pole*

In this Appendix, we prove that for continuous media invariant under the time-reversal transformation the eigenspaces at the $N^\pm$ points are equal, and hence that it is always possible to enforce the constraint (12c). Since the eigenspace at the $N^-$ point is linked with the eigenspace at the $N^+$ by the time-reversal transformation, it is sufficient to prove that the eigenspace at the $N^+$ point stays invariant under the action of $\mathcal{T}$.

As a starting point, we note that the material matrix $\mathbf{M}(\omega)$ can be decomposed as follows (see Appendix A of Ref. [22]):

$$\mathbf{M}(\omega) = \mathbf{M}_\infty + \sum_\alpha \boldsymbol{\chi}_\alpha(\omega), \tag{A1}$$

where $\mathbf{M}_\infty = \mathbf{M}(\omega = \infty)$ usually corresponds to the material matrix of the vacuum, and the generalized susceptibility $\boldsymbol{\chi}_\alpha(\omega)$ is of the form $\boldsymbol{\chi}_\alpha(\omega) = -\mathbf{B}_\alpha / (\omega - \omega_{p,\alpha})$, where

$\mathbf{B}_\alpha = -\lim_{\omega \to \omega_{p,\alpha}} \mathbf{M}(\omega)(\omega - \omega_{p,\alpha})$ and $\omega_{p,\alpha}$ is a generic pole of the material response ($\alpha = 1,2,...$). For $\omega_{p,\alpha} > 0$ the matrix $\mathbf{B}_\alpha$ must be positive semi-definite ($\mathbf{B}_\alpha \geq 0$) [22]. We also note that the electromagnetic field envelope satisfies

$$\mathbf{k} \times \mathbf{E}_\mathbf{k} = \omega\left(\frac{1}{c}\overline{\overline{\zeta}} \cdot \mathbf{E}_\mathbf{k} + \mu_0 \overline{\overline{\mu}} \cdot \mathbf{H}_\mathbf{k}\right), \qquad \mathbf{k} \times \mathbf{H}_\mathbf{k} = -\omega\left(\varepsilon_0 \overline{\overline{\varepsilon}} \cdot \mathbf{E}_\mathbf{k} + \frac{1}{c}\overline{\overline{\xi}} \cdot \mathbf{H}_\mathbf{k}\right). \qquad (A2)$$

Let us first consider an eigenspace associated with an eigenvalue $\omega_{n,k=\infty}$ such that $\omega_{n,k=\infty} \neq \omega_{p,\alpha}$ for any $\alpha$, i.e. that is not coincident with a pole of the material response. Then, in the limit $k \to \infty$ the right-hand side of the two equations in (A2) is finite because $\omega_{n,k=\infty}$ is not a pole of the material parameters. This shows that the left-hand side of the two equations also stays finite in the $k \to \infty$ limit, which is only possible if $\mathbf{E}_\mathbf{k} \sim \hat{\mathbf{k}}$ and $\mathbf{H}_\mathbf{k} \sim \hat{\mathbf{k}}$. In other words, in the $k \to \infty$ limit the modes must be longitudinal so that $\frac{1}{c}\overline{\overline{\zeta}} \cdot \mathbf{E}_\mathbf{k} + \mu_0 \overline{\overline{\mu}} \cdot \mathbf{H}_\mathbf{k} = 0$ and $\varepsilon_0 \overline{\overline{\varepsilon}} \cdot \mathbf{E}_\mathbf{k} + \frac{1}{c}\overline{\overline{\xi}} \cdot \mathbf{H}_\mathbf{k} = 0$. For continuous media with the time-reversal symmetry the permittivity and the permeability are required to be real-valued, whereas the magneto-electric tensors are pure imaginary. This establishes that the field envelope in the $k \to \infty$ limit is of the form $\mathbf{f}_\mathbf{k} = (\mathbf{E}_\mathbf{k} \quad \mathbf{H}_\mathbf{k})^T \sim (\hat{\mathbf{k}} \quad iC_{\hat{\mathbf{k}}} \hat{\mathbf{k}})^T$, where $C_{\hat{\mathbf{k}}}$ is some real-valued function that may depend on $\hat{\mathbf{k}}$. The vector $(\hat{\mathbf{k}} \quad iC_{\hat{\mathbf{k}}} \hat{\mathbf{k}})^T$ stays invariant under the action of $\mathcal{T}$. Thus, the time-reversal invariance of the material response implies that the eigenspace calculated at the $N^+$ point is also invariant under the application of $\mathcal{T}$, which concludes the proof in the case $\omega_{n,k=\infty} \neq \omega_{p,\alpha}$.

Let us now consider the case wherein $\omega_{n,k=\infty} = \omega_{p,\alpha}$, i.e. that the relevant eigenvalue is coincident with a pole of the material response. Then, writing $\mathbf{B}_\alpha = \begin{pmatrix} \mathbf{B}_\alpha^{ee} & \mathbf{B}_\alpha^{em} \\ \mathbf{B}_\alpha^{me} & \mathbf{B}_\alpha^{mm} \end{pmatrix}$ (here, all the sub-block matrices have dimensions 3×3) we see from Eq. (A1) that when $\mathbf{B}_\alpha$ is strictly positive definite the material response can be approximated by $\mathbf{M} \approx -\dfrac{1}{\omega - \omega_{p,\alpha}} \begin{pmatrix} \mathbf{B}_\alpha^{ee} & \mathbf{B}_\alpha^{em} \\ \mathbf{B}_\alpha^{me} & \mathbf{B}_\alpha^{mm} \end{pmatrix}$. Then, in the $k \to \infty$ limit (and for $\omega$ in the vicinity of $\omega_{p,\alpha}$) Eq. (A2) reduces to

$$\hat{\mathbf{k}} \times \mathbf{E}_\mathbf{k} = -\frac{\omega_{p,\alpha}}{k\Delta\omega}\left(\mathbf{B}_\alpha^{me} \cdot \mathbf{E}_\mathbf{k} + \mathbf{B}_\alpha^{mm} \cdot \mathbf{H}_\mathbf{k}\right), \quad \hat{\mathbf{k}} \times \mathbf{H}_\mathbf{k} = \frac{\omega_{p,\alpha}}{k\Delta\omega}\left(\mathbf{B}_\alpha^{ee} \cdot \mathbf{E}_\mathbf{k} + \mathbf{B}_\alpha^{em} \cdot \mathbf{H}_\mathbf{k}\right), \quad (A3)$$

where $\Delta\omega = \omega - \omega_{p,\alpha}$. Noting that $\mathbf{B}_\alpha^{ee}, \mathbf{B}_\alpha^{mm}$ must be real-valued and that $\mathbf{B}_\alpha^{em}, \mathbf{B}_\alpha^{em}$ must be pure imaginary it is simple to verify that the $\mathcal{T} = \mathcal{K}\mathcal{U}$ operator transforms (without flipping the wave vector) solutions $\mathbf{f}_\mathbf{k} = (\mathbf{E}_\mathbf{k} \quad \mathbf{H}_\mathbf{k})^T$ of Eq. (A3) associated with some (nonzero) eigenvalue $\tilde{\lambda} = \dfrac{\omega_{p,\alpha}}{k\Delta\omega}$ into solutions $\mathcal{T} \cdot \mathbf{f}_\mathbf{k}$ associated with the symmetric eigenvalue $-\dfrac{\omega_{p,\alpha}}{k\Delta\omega}$. These two branches of eigenmodes (with asymptotic dispersion $\omega \approx \omega_{p,\alpha} \pm \dfrac{\omega_{p,\alpha}}{k\tilde{\lambda}}$) touch in the $k \to \infty$ limit, and hence the eigenspace associated with the $N^+$ point is invariant under the action of the $\mathcal{T}$ operator. This implies the eigenspaces at the $N^\pm$ points are indeed equal as we wanted to prove.

It is emphasized that the approximation $\mathbf{M} \approx -\dfrac{1}{\omega - \omega_{p,\alpha}} \begin{pmatrix} \mathbf{B}_\alpha^{ee} & \mathbf{B}_\alpha^{em} \\ \mathbf{B}_\alpha^{me} & \mathbf{B}_\alpha^{mm} \end{pmatrix}$ requires that $\mathbf{B}_\alpha$ is a positive definite matrix. However, the proof can be generalized to the case wherein $\mathbf{B}_\alpha \geq 0$. To illustrate this, suppose that the resonance associated with $\omega_{p,\alpha}$ has a purely electric nature, so that $\mathbf{B}_\alpha^{em} = \mathbf{B}_\alpha^{em} = 0$ and $\mathbf{B}_\alpha^{mm} = 0$. In this case, Eq. (A3) needs to be replaced by

$$\mathbf{H}_\mathbf{k} = \bar{\mu}^{-1}(\omega_{p,\alpha}) \cdot \left[ \frac{1}{\omega_{p,\alpha} \mu_0} \mathbf{k} \times \mathbf{E}_\mathbf{k} - \frac{1}{\mu_0 c} \bar{\zeta}(\omega_{p,\alpha}) \cdot \mathbf{E}_\mathbf{k} \right], \quad \hat{\mathbf{k}} \times \mathbf{H}_\mathbf{k} = \frac{\omega_{p,\alpha}}{k \Delta \omega} \mathbf{B}_\alpha^{ee} \cdot \mathbf{E}_\mathbf{k}. \quad (A4)$$

Hence, in the $k \to \infty$ limit it is possible to write $\mathbf{H}_\mathbf{k} \approx \dfrac{1}{\omega_{p,\alpha} \mu_0} \bar{\mu}^{-1}(\omega_{p,\alpha}) \cdot (\mathbf{k} \times \mathbf{E}_\mathbf{k})$ so that

$\hat{\mathbf{k}} \times \left[ \bar{\mu}^{-1}(\omega_{p,\alpha}) \cdot (\hat{\mathbf{k}} \times \mathbf{E}_\mathbf{k}) \right] = \dfrac{\mu_0 \omega_{p,\alpha}^2}{k^2 \Delta \omega} \mathbf{B}_\alpha^{ee} \cdot \mathbf{E}_\mathbf{k}$. Therefore, in the $k \to \infty$ limit the magnetic field is dominant over the electric field, and the eigenspace at the $N^+$ point is generated by a vector of the form $\mathbf{f}_\mathbf{k} = (0 \quad \mathbf{H}_\mathbf{k})^T$ with $\mathbf{H}_\mathbf{k}$ real-valued. Thus, the eigenspace at the $N^+$ point is again invariant under the action of $\mathcal{T}$, and hence it must be coincident with the eigenspace at the $N^-$ point.

## Appendix B: The $\mathbb{Z}_2$ index definition in electronics and in photonics

It is interesting to contrast the definition of the $\mathbb{Z}_2$ number in Eq. (13) with that of Fu and Kane for electronic systems $\tilde{\mathcal{D}} = \mathcal{D}_{EBZ} \bmod 2$ [21]. As seen, in our definition for photonic systems there is an extra factor of 2 before the $\mathcal{D}_{EBZ}$ symbol. Interestingly, it is possible to reconcile the two definitions by noting that our basis $\mathbf{f}_{n\mathbf{k}}$ only includes the *positive frequency* eigenfunctions. Notably, it is possible to pair each positive frequency eigenmode ($\mathbf{f}_{n\mathbf{k}}$) with a negative frequency partner ($\mathbf{f}_{n\mathbf{k}}^-$) through the mapping $\mathbf{f}_{n\mathbf{k}} \to \mathbf{f}_{n\mathbf{k}}^- = \mathcal{U} \cdot \mathbf{f}_{n\mathbf{k}}$. Indeed, our system is invariant under the action of $\mathcal{U} = \mathcal{K}\mathcal{T}$ because of the invariance of the material response to both $\mathcal{K}$ (reality condition, satisfied by any photonic system) and $\mathcal{T}$ (time reversal). Note that the operator $\mathcal{U}$ flips the frequency without changing the wave vector ($\omega_{n\mathbf{k}}^- = -\omega_{n\mathbf{k}}$), and that $\mathcal{U}$ is idempotent, i.e. $\mathcal{U}^2 = \mathbf{1}$. The Berry potential associated with the negative frequency branches (defined as $\mathbf{f}_{n\mathbf{k}}^- = \mathcal{U} \cdot \mathbf{f}_{n\mathbf{k}}$) satisfies $\mathcal{A}_\mathbf{k}^- = \mathcal{A}_\mathbf{k}$. This can be proven using $\mathbf{f}_{n\mathbf{k}}^- = \mathcal{U} \cdot \mathbf{f}_{n\mathbf{k}}$ and the invariance of the material response under the operator $\mathcal{U}$ (which implies that $\mathbf{M}(-\omega) = \mathcal{U} \cdot \mathbf{M}(\omega) \cdot \mathcal{U}$) in the definition of the Berry potential [Eq. (3)]. Moreover, the negative frequency branches satisfy the same Gauge restrictions [Eq. (12)] as the positive frequency branches. Thus, if one includes both positive and negative frequency branches in the calculation of the Berry potential (so that the relevant spectral range becomes $\omega_{min} < |\omega| < \omega_{max}$), the topological index $\mathcal{D}$ needs to be redefined as $\mathcal{D}_{EBZ} \bmod 2$ exactly as in electronics [21]. Indeed, the contribution of the negative frequency branches is identical to that of the positive frequency branches.

The previous discussion also reveals that the pairing of positive frequency and negative frequency eigenmodes has some analogies with the pairing between Kramers partners in electronics (the Kramers pairs are however linked by the time reversal transformation). This analogy is particularly meaningful when the positive and negative frequency branches are entangled, i.e. for $\omega = 0$ and $\omega = \infty$.

## *Appendix C: Proof that the $\mathcal{D}$ number is an integer*

Here, we demonstrate that the $\mathcal{D}$ number is an integer. We start by noting that from Eq. (14) it is possible to write (supposing that the *EBZ* is the region $k_y > 0$):

$$\mathcal{D} = \left[ \frac{1}{\pi} \int_0^\pi \left( \mathcal{A}_\varphi k \right) \Big|_{k=0^+} d\varphi - \frac{1}{\pi} \int_0^\pi \left( \mathcal{A}_\varphi k \right) \Big|_{k=\infty^+} d\varphi \right] \mod 2, \quad (C1)$$

where $(k, \varphi)$ determine a system of polar coordinates centered at the origin of the **k**-space and $\mathcal{A}_\varphi = \mathcal{A}_\mathbf{k} \cdot \hat{\boldsymbol{\varphi}}$. From Eq. (3) the Berry potential associated with the *n*-th band satisfies:

$$k\mathcal{A}_{n,\varphi} = \frac{\mathrm{Re}\left\{ i \mathbf{f}_{n,k\varphi}^* \cdot \frac{\partial}{\partial \omega}\left[ \omega \mathbf{M}(\omega) \right]_{\omega_{n\mathbf{k}}} \cdot \partial_\varphi \mathbf{f}_{n,k\varphi} \right\}}{\mathbf{f}_{n,k\varphi}^* \cdot \frac{\partial}{\partial \omega}\left[ \omega \mathbf{M}(\omega) \right]_{\omega_{n\mathbf{k}}} \cdot \mathbf{f}_{n,k\varphi}}, \quad (C2)$$

where $\mathbf{f}_{n,k\varphi} = \mathbf{f}_{n,\mathbf{k}=k\hat{\boldsymbol{\varphi}}}$ and $\partial_\varphi = \partial/\partial \varphi$. In the following subsections it is shown that both $\frac{1}{\pi} \int_0^\pi \left( \mathcal{A}_\varphi k \right) \Big|_{k=0^+} d\varphi$ (contribution from the south-pole) and $\frac{1}{\pi} \int_0^\pi \left( \mathcal{A}_\varphi k \right) \Big|_{k=\infty^+} d\varphi$ (contribution from the north-pole) are integer numbers.

*I. Contribution from the south-pole*

To prove that the contribution from the south-pole is an integer, we use the fact that the pseudo-Hamiltonian that characterizes the material response is smooth for any finite **k** and hence the eigenspaces must vary smoothly with **k** [22]. Thus, for a subset of bands such that the positive frequency branches are disconnected from the negative frequency branches at the origin ($\omega_{n,k=0^+} \neq 0$) it is possible to write:

$$\left( \mathbf{f}_{n,\varphi} \right) = \mathbf{V}_\varphi \cdot \left( \mathbf{f}_{n0} \right), \quad (C3)$$

where $\mathbf{f}_{n,\varphi} = \mathbf{f}_{n,k\varphi}\big|_{k=0^+}$, $\mathbf{f}_{n0} = \mathbf{f}_{n,k\varphi}\big|_{k=0^+,\varphi=0^+}$, and $\mathbf{V}_\varphi$ is some unitary matrix that varies continuously with $\varphi$. Hence, the Berry potential associated with this subset of bands satisfies $\left(\mathcal{A}_\varphi k\right)\big|_{k=0^+} = i\partial_\varphi \left[\ln \det \mathbf{V}_\varphi\right]$ so that:

$$\frac{1}{\pi}\int_0^\pi \left(\mathcal{A}_\varphi k\right)\big|_{k=0^+} d\varphi = \frac{i}{\pi}\left(\ln \det \mathbf{V}_{\varphi=\pi} - \ln \det \mathbf{V}_{\varphi=0}\right). \tag{C4}$$

It is seen from the definition that $\mathbf{V}_{\varphi=0} = \mathbf{1}$ is the identity matrix. On the other hand, from the Gauge constraint (12a) it is necessary that $\left(\mathbf{f}_{n\mathbf{k}}\right)_{S^-} = \tilde{\mathbf{V}} \cdot \left(\mathbf{f}_{n\mathbf{k}}\right)_{S^+}$ with $\tilde{\mathbf{V}}$ a unitary transformation with $\det \tilde{\mathbf{V}} = 1$. Clearly, this requires that $\det\left(\mathbf{V}_{\varphi=\pi}\right) = 1$. Thus, $\ln \det \mathbf{V}_{\varphi=\pi} - \ln \det \mathbf{V}_{\varphi=0} = i2\pi l$, where $l$ is some integer number. This demonstrates that the eigenfunction branches with $\omega_{n,k=0^+} \neq 0$ give an integer contribution to the first term in the right-hand side of Eq. (C1). Moreover, the contribution of these terms is trivial because the previous analysis also shows that the right-hand side of Eq. (C4) is an even integer.

Let us now consider the branches for which $\omega_{n,k=0^+} = 0^+$. For these branches, in general it is not possible to write $\left(\mathbf{f}_{n,\varphi}\right) = \mathbf{V}_\varphi \cdot \left(\mathbf{f}_{n0}\right)$ because the positive frequency branches are linked with the negative frequency branches and hence the respective eigenspaces are entangled. In other words, the eigenspaces generated by $\left(\mathbf{f}_{n,\varphi}\right)$ typically depend on $\varphi$. As discussed in Sect. II.C, in the long wavelength limit ($\omega \to 0$) the material response $\mathbf{M}(\omega=0)$ is real-valued. This implies that the $\varphi$-dependent eigenspaces must be invariant under the operation of complex conjugation, i.e. in the $\omega \to 0$ limit it is possible to pick a basis $\left( \tilde{\mathbf{f}}_{n,\varphi} \right)$ of real-valued vectors for each eigenspace. Note that $\left( \tilde{\mathbf{f}}_{n,\varphi} \right)$ may not be eigenvectors, they are only required to be real-valued and to generate the relevant eigenspace. Clearly, now we can write $\left( \mathbf{f}_{n,\varphi} \right) = \mathbf{V}_\varphi \cdot \left( \tilde{\mathbf{f}}_{n,\varphi} \right)$ for some unitary matrix $\mathbf{V}_\varphi$. Because $\left( \tilde{\mathbf{f}}_{n,\varphi} \right)$ are real valued it is simple to check that the corresponding Berry potential vanishes. Hence, the Berry potential associated with $\left( \mathbf{f}_{n,\varphi} \right)$ also satisfies $\left( \mathcal{A}_\varphi k \right)\big|_{k=0^+} = i\partial_\varphi \left[ \ln \det \mathbf{V}_\varphi \right]$, and thus Eq. (C4) also holds for bands with $\omega_{n,k=0^+} = 0^+$.

Next, we note that because of the time-reversal invariance there is a unitary matrix $\mathbf{W}$ such that $\left( \mathcal{T} \tilde{\mathbf{f}}_{n,\varphi=0} \right) = \mathbf{W} \cdot \tilde{\mathbf{f}}_{n,\varphi=\pi}$. But since by hypothesis $\left( \tilde{\mathbf{f}}_{n,\varphi} \right)$ is real-valued it follows that $\left( \mathcal{U} \cdot \tilde{\mathbf{f}}_{n,\varphi=0} \right) = \mathbf{W} \cdot \tilde{\mathbf{f}}_{n,\varphi=\pi}$. Therefore the unitary matrix $\mathbf{W}$ must also be real-valued, and hence it must be such that $\det \mathbf{W} = \pm 1$.

Using $\left( \mathbf{f}_{n,\varphi} \right) = \mathbf{V}_\varphi \cdot \left( \tilde{\mathbf{f}}_{n,\varphi} \right)$ and $\left( \mathcal{U} \cdot \tilde{\mathbf{f}}_{n,\varphi=0} \right) = \mathbf{W} \cdot \tilde{\mathbf{f}}_{n,\varphi=\pi}$ it is simple to check that $\left( \mathbf{f}_{n,\varphi=0} \right) = \mathbf{V}_{\varphi=0} \cdot \mathbf{W} \cdot \mathbf{V}_{\varphi=\pi}^{-1} \left( \mathcal{U} \cdot \mathbf{f}_{n,\varphi=\pi} \right)$. But the Gauge restrictions [Eq. (12b)] now imply that $\left( \mathbf{f}_{n\mathbf{k}} \right)_{S^-} = \tilde{\mathbf{V}} \cdot \left( \mathcal{U} \cdot \mathbf{f}_{n\mathbf{k}} \right)_{S^+}$ where $\tilde{\mathbf{V}}$ is some unitary transformation with $\det \tilde{\mathbf{V}} = 1$. This is only possible if $\det \left( \mathbf{V}_{\varphi=0} \cdot \mathbf{W} \cdot \mathbf{V}_{\varphi=\pi}^{-1} \right) = 1$, which establishes that $\ln \det \mathbf{V}_{\varphi=\pi} - \ln \det \mathbf{V}_{\varphi=0} = i2\pi l + \ln \det \mathbf{W}$, where $l$ is an integer. Then, because $\det \mathbf{W} = \pm 1$ the right-hand side of Eq. (C4) is indeed an integer, which proves the desired result.

In summary, the contribution of the south-pole to the topological invariant is trivial except possibly for eigenmode branches with $\omega_{n,k=0^+} = 0^+$.

*II. Contribution from the north-pole for bands with $\omega_{n,k=\infty} = +\infty$*

Next, we calculate the contribution to the $\mathcal{D}$ number (north-pole contribution) from the bands with $\omega_{n,k=\infty} = +\infty$. In this case, we define $\mathbf{f}_{n,\varphi} = \mathbf{f}_{n,k\varphi}\big|_{k=+\infty}$ and use the fact that in the high-frequency limit the material response is asymptotically the same as that of the vacuum. Hence, $\mathbf{M}(\omega = \infty)$ can be assumed real-valued. This allows us to write $(\mathbf{f}_{n,\varphi}) = \mathbf{V}_\varphi \cdot (\tilde{\mathbf{f}}_{n,\varphi})$ for some unitary matrix $\mathbf{V}_\varphi$ and for some family of vectors $(\tilde{\mathbf{f}}_{n,\varphi})$ real-valued. Hence, by slightly modifying the arguments used in the subsection *I* of this Appendix for the case $\omega_{n,k=0^+} = 0$ it is possible to prove that the bands with $\omega_{n,k=\infty} = +\infty$ yield an integer contribution to the second integral in the right-hand side of Eq. (C1).

*III. Contribution from the north-pole for bands with $\omega_{n,k=\infty} \neq \infty$*

Finally, we analyze contributions to the $\mathcal{D}$ number arising from eigenmode branches with $\omega_{n,k=\infty}$ finite. As in the previous subsection, we put $\mathbf{f}_{n,\varphi} = \mathbf{f}_{n,k\varphi}\big|_{k=+\infty}$.

First we consider the case wherein $\omega_{n,k=\infty} \neq \omega_{p,\alpha}$ for any $\alpha$, i.e. the pertinent eigenvalue differs from the poles of the material response (see Appendix A). It was proven in Appendix A that in the $k \to \infty$ limit the eigenspace is generated by a vector of the form $\tilde{\mathbf{f}}_{n,\varphi} = \begin{pmatrix} \hat{\mathbf{k}} & iC_{n,\hat{\mathbf{k}}} \hat{\mathbf{k}} \end{pmatrix}^T$, where $C_{n,\hat{\mathbf{k}}}$ is some real-valued constant that may depend on $\hat{\mathbf{k}}$. Importantly, the time-reversal invariance implies that $C_{n,\hat{\mathbf{k}}} = C_{n,-\hat{\mathbf{k}}}$. Assuming for simplicity that $\omega_{n,k=\infty}$ is a non-degenerate eigenvalue, it is then possible to write $\mathbf{f}_{n,\varphi} = e^{i\theta_\varphi} \tilde{\mathbf{f}}_{n,\varphi}$. The Berry potential associated with $\tilde{\mathbf{f}}_{n,\varphi}$ vanishes. Thus, the Berry potential associated $(\mathbf{f}_{n,\varphi})$ satisfies $(\mathcal{A}_\varphi k)\big|_{k=\infty^+} = -\partial_\varphi \theta_\varphi$. Hence, it follows that

$$\frac{1}{\pi}\int_0^\pi (\mathcal{A}_\varphi k)\big|_{k=\infty^+} d\varphi = \frac{-1}{\pi}(\theta_{\varphi=\pi} - \theta_{\varphi=0}). \tag{C5}$$

Using the fact that $C_{n,\hat{\mathbf{k}}} = C_{n,-\hat{\mathbf{k}}}$ we see that $(\tilde{\mathbf{f}}_{n,\varphi})_{N^-} = -(\tilde{\mathbf{f}}_{n,\varphi})_{N^+}$. Thus, to ensure the Gauge restriction (12c) it is necessary that $\theta_{\varphi=\pi} - \theta_{\varphi=0} = \pi + 2\pi l$, where $l$ is an integer. This implies that the right-hand side of Eq. (C5) is an odd integer. Thus, we conclude that each non-degenerate band with $\omega_{n,k=\infty} \neq \omega_{p,\alpha}$ gives a contribution +1 to the $\mathcal{D}$ number.

The case wherein $\omega_{n,k=\infty} = \omega_{p,\alpha}$ with $\mathbf{B}_\alpha$ a positive semi-definite matrix can be treated using similar arguments (see the definition of $\mathbf{B}_\alpha$ in Appendix A). To illustrate this, we consider the particular case of a purely electric resonance for which $\mathbf{B}_\alpha^{em} = \mathbf{B}_\alpha^{em} = 0$ and $\mathbf{B}_\alpha^{mm} = 0$. As shown in Appendix A, for a purely electric resonance the relevant eigenspace in the $k \to \infty$ limit is generated by vectors of the form $\tilde{\mathbf{f}}_{n,\varphi} = \begin{pmatrix} \mathbf{0} & \mathbf{H}_{n,\mathbf{k}} \end{pmatrix}^T$ with $\mathbf{H}_{n,\mathbf{k}}$ real-valued functions. It is evidently possible to link $(\mathbf{f}_{n,\varphi})$ with $(\tilde{\mathbf{f}}_{n,\varphi})$ by a unitary transformation $\mathbf{V}_\varphi$ such that $(\mathbf{f}_{n,\varphi}) = \mathbf{V}_\varphi \cdot (\tilde{\mathbf{f}}_{n,\varphi})$. Since the Berry potential associated with $(\tilde{\mathbf{f}}_{n,\varphi})$ vanishes, we find that the Berry potential associated with $(\mathbf{f}_{n,\varphi})$ satisfies $(\mathcal{A}_\varphi k)\big|_{k=\infty^+} = i\partial_\varphi [\ln \det \mathbf{V}_\varphi]$ so that:

$$\frac{1}{\pi}\int_0^\pi (\mathcal{A}_\varphi k)\big|_{k=\infty^+} d\varphi = \frac{i}{\pi}(\ln \det \mathbf{V}_{\varphi=\pi} - \ln \det \mathbf{V}_{\varphi=0}). \tag{C6}$$

Because of the time reversal invariance it is possible to write $(T\tilde{\mathbf{f}}_{n,\varphi=0}) = \mathbf{W} \cdot \tilde{\mathbf{f}}_{n,\varphi=\pi}$ for some unitary matrix $\mathbf{W}$. The matrix $\mathbf{W}$ must be real-valued because $(\tilde{\mathbf{f}}_{n,\varphi})$ also is, and hence $\det(\mathbf{W}) = \pm 1$. Moreover, because $\tilde{\mathbf{f}}_{n,\varphi}$ is of the form $\tilde{\mathbf{f}}_{n,\varphi} = \begin{pmatrix} \mathbf{0} & \mathbf{H}_{n,\mathbf{k}} \end{pmatrix}^T$ it is obvious that $T\tilde{\mathbf{f}}_{n,\varphi=0} = -\tilde{\mathbf{f}}_{n,\varphi=0}$. Therefore, we conclude that $(\tilde{\mathbf{f}}_{n,\varphi=0}) = -\mathbf{W} \cdot \tilde{\mathbf{f}}_{n,\varphi=\pi}$.

Using now $(\mathbf{f}_{n,\varphi}) = \mathbf{V}_\varphi \cdot (\tilde{\mathbf{f}}_{n,\varphi})$ and $(\tilde{\mathbf{f}}_{n,\varphi=0}) = -\mathbf{W} \cdot \tilde{\mathbf{f}}_{n,\varphi=\pi}$ one sees that $(\mathbf{f}_{n,\varphi=0}) = -\mathbf{V}_{\varphi=0} \cdot \mathbf{W} \cdot \mathbf{V}_{\varphi=\pi}^{-1} (\mathbf{f}_{n,\varphi=\pi})$. But the Gauge constraint (12c) forces that $\det(-\mathbf{V}_{\varphi=0} \cdot \mathbf{W} \cdot \mathbf{V}_{\varphi=\pi}^{-1}) = 1$. Taking into account that $\det(\mathbf{W}) = \pm 1$ we see that this implies that $\dfrac{\det(\mathbf{V}_{\varphi=0})}{\det(\mathbf{V}_{\varphi=\pi})} = \pm 1$, and thus the right-hand side of Eq. (C6) is really an integer number.

Let us finally consider the case wherein the relevant eigenspace is associated with a pole of the material response, $\omega_{n,k=\infty} = \omega_{p,\alpha}$, for which $\mathbf{B}_\alpha$ is strictly positive definite. It is clear from the discussion of Appendix A that in this case it is possible to split the eigenmode branches into two subfamilies: the branches $\mathbf{f}_{n\mathbf{k}}^U$ for which $\omega_{n,k=\infty}^U = \omega_{p,\alpha} + 0^+$ and the branches $\mathbf{f}_{n\mathbf{k}}^D$ for which $\omega_{n,k=\infty}^D = \omega_{p,\alpha} - 0^+$. In other words, the family associated with superscript $U$ (up) approaches the pole with $\omega_{n,\mathbf{k}}^U > \omega_{p,\alpha}$ (in the limit $k \to \infty$), whereas the family associated with the superscript $D$ (down) approaches the pole with $\omega_{n,\mathbf{k}}^D < \omega_{p,\alpha}$. The analysis of the Appendix A also shows that in the $k \to \infty$ limit the space generated by $(\mathbf{f}_{n\mathbf{k}}^U)$ is linked to the space generated by $(\mathbf{f}_{n\mathbf{k}}^D)$ by the time reversal operator $\mathcal{T}$. Thus, there is a unitary matrix $\mathbf{V}_\varphi$ such that $(\mathbf{f}_{n,\varphi}^D) = \mathbf{V}_\varphi (\mathcal{T} \cdot \mathbf{f}_{n,\varphi}^U)$ where $\mathbf{f}_{n,\varphi}^U = \mathbf{f}_{n,\mathbf{k}=k\hat{\boldsymbol{\varphi}}}^U \big|_{k=+\infty}$, etc.

To proceed, we note that the time-reversal invariance implies that $\mathbf{M}(\omega) = \mathcal{U} \cdot \mathbf{M}^*(\omega) \cdot \mathcal{U}$ (see Sect. II.A) and because of this

$$\mathrm{Re}\left\{i\mathcal{T}\mathbf{f}_{n,\varphi}^* \cdot \frac{\partial}{\partial\omega}[\omega\mathbf{M}(\omega)]_{\omega_{n\mathbf{k}}} \cdot \partial_\varphi \mathcal{T}\mathbf{f}_{n,\varphi}\right\} = \mathrm{Re}\left\{i\mathbf{f}_{n,\varphi} \cdot \frac{\partial}{\partial\omega}[\omega\mathcal{U} \cdot \mathbf{M}(\omega) \cdot \mathcal{U}]_{\omega_{n\mathbf{k}}} \cdot \partial_\varphi \mathbf{f}_{n,\varphi}^*\right\}$$

$$= -\mathrm{Re}\left\{i\mathbf{f}_{n,\varphi}^* \cdot \frac{\partial}{\partial\omega}[\omega\mathcal{U} \cdot \mathbf{M}^*(\omega) \cdot \mathcal{U}]_{\omega_{n\mathbf{k}}} \cdot \partial_\varphi \mathbf{f}_{n,\varphi}\right\}.$$

$$= -\mathrm{Re}\left\{i\mathbf{f}_{n,\varphi}^* \cdot \frac{\partial}{\partial\omega}[\omega\mathbf{M}(\omega)]_{\omega_{n\mathbf{k}}} \cdot \partial_\varphi \mathbf{f}_{n,\varphi}\right\}$$

(C7)

Hence, from Eq. (C2) one sees that the Berry potentials associated with $(\mathbf{f}_{n,\varphi}^U)$ and $(\mathcal{T} \cdot \mathbf{f}_{n,\varphi}^U)$ cancel out. This shows that the Berry potential determined by $(\mathbf{f}_{n\mathbf{k}}^U)$ and $(\mathbf{f}_{n\mathbf{k}}^D)$ satisfies $(\mathcal{A}_\varphi k)\big|_{k=\infty^+} = i\partial_\varphi [\ln \det \mathbf{V}_\varphi]$ so that Eq. (C6) also applies in this case. On the other hand, because of the time-reversal invariance there is a unitary matrix $\mathbf{W}$ such that $(\mathcal{T} \cdot \mathbf{f}_{n,\varphi=0}^U) = \mathbf{W} \cdot (\mathbf{f}_{n,\varphi=\pi}^U)$. It is simple to check that this relation together with $(\mathbf{f}_{n,\varphi}^D) = \mathbf{V}_\varphi (\mathcal{T} \cdot \mathbf{f}_{n,\varphi}^U)$ give:

$$\begin{pmatrix} \mathbf{f}_{n,\varphi=0}^U \\ \mathbf{f}_{n,\varphi=0}^D \end{pmatrix} = \underbrace{\begin{pmatrix} \mathbf{0} & \mathbf{W}^* \cdot \mathbf{V}_{\varphi=\pi}^{-1} \\ \mathbf{V}_{\varphi=0} \cdot \mathbf{W} & \mathbf{0} \end{pmatrix}}_{\tilde{\mathbf{V}}} \begin{pmatrix} \mathbf{f}_{n,\varphi=\pi}^U \\ \mathbf{f}_{n,\varphi=\pi}^D \end{pmatrix}. \qquad (C8)$$

The Gauge restriction (12c) forces the matrix $\tilde{\mathbf{V}}$ to have determinant +1. This is only possible if $\det(\mathbf{W}^* \cdot \mathbf{V}_{\varphi=\pi}^{-1}) \det(\mathbf{V}_{\varphi=0} \cdot \mathbf{W}) = (-1)^N$ where $N$ is the dimension of the matrix $\mathbf{V}_\varphi$, i.e. the number of elements of the subfamily $(\mathbf{f}_{n\mathbf{k}}^U)$. Because $\mathbf{W}$ is a unitary matrix it follows that $\det(\mathbf{W}^*) = \det(\mathbf{W})^{-1}$ and thus we conclude that $\frac{\det(\mathbf{V}_{\varphi=0})}{\det(\mathbf{V}_{\varphi=\pi})} = (-1)^N$. This result confirms that the right-hand side of Eq. (C6) is indeed an integer. Moreover, the contribution of the considered eigenspace to the invariant is trivial for *N* even and is +1 for *N* odd.

[33] K. Y. Bliokh, F. J. Rodríguez-Fortuño, F. Nori, and A.V. Zayats, "Spin-orbit interactions of light", *Nat. Photonics* **9**, 796 (2015).

[34] Supplementary online materials with the field animations of Figs. 4a and 4b.

[35] D. Colton, R. Kress, *Inverse Acoustic and Electromagnetic Scattering Theory*, (2$^{nd}$ Ed.), Springer, Berlin, 1998, p.69.

[36] Shun-Qing Shen, *Topological Insulators,* (Series in Solid State Sciences,174), Springer, Berlin, 2012 (p. 98).

[37] D. M. Pozar, *Microwave Engineering,* 4$^{th}$ Ed. Wiley, Hoboken, NJ, 2012.

[38] M. M. I. Saadoun, N. Engheta, "A reciprocal phase shifter using novel pseudochiral or Ω medium", *Microwave and Opt. Tech. Lett.* **5** (4), 184, (1992).

[39] S. A. Tretyakov, A. A. Sochava, "Novel uniaxial bianisotropic materials: reflection and transmission in planar structures", *Progress in Electromagnetic Research* (PIER), **9**, 157, (1994).